\documentclass[aip,jcp,preprint,floatfix,longbibliography]{revtex4-1}

\usepackage{graphicx}
\usepackage{amsmath}
\usepackage{braket}
\usepackage{bm}
\usepackage{siunitx}

\newcommand{\ie}{i.e.}

\newcommand{\rmd}{d}

\newcommand{\rmi}{i}

\begin{document}

\title{Rotational dynamics of a diatomic molecular ion in a Paul trap}

\author{A. Hashemloo}
\affiliation{Department of Physics, Ume{\aa} University, SE-901\,87 Ume{\aa},
  Sweden}

\author{C. M. Dion}
\email{claude.dion@umu.se}
\affiliation{Department of Physics, Ume{\aa} University, SE-901\,87 Ume{\aa},
  Sweden}

\begin{abstract}
  We present models for a heteronuclear diatomic molecular ion in a
  linear Paul trap in a rigid-rotor approximation, one purely
  classical, the other where the center-of-mass motion is treated
  classically while rotational motion is quantized.  We study the
  rotational dynamics and their influence on the motion of the
  center-of-mass, in the presence of the coupling between the
  permanent dipole moment of the ion and the trapping electric field.
  We show that the presence of the permanent dipole moment affects
  the trajectory of the ion, and that it departs from the Mathieu
  equation solution found for atomic ions.  For the case of quantum
  rotations, we also evidence the effect of the above-mentioned
  coupling on the rotational states of the ion.
\end{abstract}

\date{\today}

\maketitle

\section{Introduction}
\label{sec:intro}

Laser-cooled or sympathetically-cooled atomic and molecular ions have
been in the spotlight in recent years, due to their great potential in the
study and development of many fields, such as chemical
reactions,\cite{Drewsen_2000_PR, Blythe_2005_PR, Baba_2002_JCP, %
 Tong_CPL_2012, Chang_Science_2013, Rosch_JCP_2014}
high-precision spectroscopy,\cite{Douglas_MSR_2005,%
  Leanhardt_JMS_2011, Loh_Science_2013}
atomic optics, and in many other fast developing fields
related to quantum information and quantum computing, where the
manipulation of the internal states of these ions has become
possible.\cite{Weidinger_CP_2008, Buluta_2008_APS, Alonso_NJP_2013,%
  Amniat-Talab_EPJD_2012, Leibfried_NJP_2012, Shi_NJP_2013}

One of the most applied techniques to trap these cold ions is known as
the Paul trap, using time-dependent radio-frequency electric
fields.\cite{Paul_RMP_1990} Paul traps, specifically those with a
linear configuration,\cite{Prestage_JAP_1989} consist of electrodes
designed to surround an inner space where the trapping of a single
charged particle, as well as the simultaneous trapping a string of
charged particles, occurs. This makes them highly suitable to be used
in experiments involving laser control of the ions, since the spacing
between the electrodes as well as the spacing between the ions provide
a good environment for this
purpose.\cite{Raizen_PRA_1992,Raizen_JMO_1992,Brkic_PRA_2006} In this
regard, large scale studies of trapped ions in large scales, in the
form of Coulomb crystals, have become feasible, with promising
developments for large-scale quantum simulations and quantum
computations.\cite{Buluta_2008_APS, Drewsen_2003_MS, Porras_2006_PR,%
  Willitsch_IRPC_2012}

While there have been many experimental studies using molecular ions,
on the theoretical side there have been few attempts to treat them
differently than atomic ions, at least with respect to their
interaction with the trapping electric field.  In particular, the
rotation of the molecular ions has been considered mostly with respect
to a coupling with an external laser field.\cite{Vogelius_JPB_2006,
  Lazarou_PRA_2010, Shi_NJP_2013,%
  Berglund_2015_NJP}

In this paper we study the dynamical behavior of a rigid,
heteronuclear diatomic molecular ion trapped inside a linear Paul
trap. We investigate the coupling of the permanent dipole moment of
the molecular ion with the trapping electric field and its effect on
the rotational dynamics of the ion and also on the center-of-mass
motion of the ion. We develop classical and semi-classical models of
the molecular ion, where rotation is treated classically and quantum
mechanically, respectively.  We carry out simulations of the motion of
the center of mass, including the effects of the interaction between
the trapping electric field and the permanent dipole moment of the
molecular ion, by solving the classical equations of motion.
Classical rotation is treated using the ``body vector
method,''\cite{evans1977representatation} while a basis set of
spherical harmonics is used to describe the quantum rotational wave
function.

This paper is organised as follows. First, we introduce our model for
a rigid diatomic molecular ion trapped inside a linear Paul trap
(Sec.~\ref{sec:hamil}). The numerical methods for the classical and
semi-classical simulations are described in Sec.~\ref{sec:num}. The
results obtained from both classical and semi-classical methods are
presented in Sec.~\ref{subsec:res_class} and
Sec.~\ref{subsec:res_semi_class}, respectively. Finally, concluding
remarks are given in Sec.~\ref{sec:con}.

\section{Hamiltonian of a rigid diatomic molecular ion in a linear
  Paul trap}
\label{sec:hamil}

\subsection{Quantum-mechanical Hamiltonian}
\label{subsec:QMH}

We consider a diatomic molecular ion inside a linear Paul
trap,\cite{Prestage_JAP_1989,Major_book_2005} combining a quadrupole,
time-dependent, radio-frequency electric potential and a static
electric potential,
\begin{equation}
\label{eq:tr_pot}
\Phi(t) = \Phi_{\mathrm {rf}}(t) +\Phi_{\mathrm s}\, ,
\end{equation}
in which, for an ion located at position $(X,Y,Z)$,
\begin{subequations}
\label{eq:tr_pots}
\begin{align}
\Phi_{\mathrm {rf}}(t) &= \frac{1}{2} \frac{V_{0}}{r_{0}^{2}}\left ( X^{2} - Y^{2} \right ) \cos \Omega t\, ,\\
\Phi_{\mathrm s} &= \frac{\kappa U_{0}}{z_{0}^{2}}\left [ Z^{2} - \frac{1}{2}\left ( X^{2} + Y^{2} \right ) \right ]\, ,
\end{align}
\end{subequations}
where $V_{0}$ and $U_{0}$ are the amplitudes of the radio-frequency
(with frequency $\Omega$) and static potentials, respectively, $r_{0}$
and $z_{0}$ are parameters representing the dimensions of the trap, and
$\kappa$ is the geometric factor, which is solely dependent on the
configuration of the trap.\cite{Raizen_PRA_1992}

The Hamiltonian of a free diatomic molecular ion in a \textit{rigid
  rotor} model, where the nuclear vibrations of the ion are ignored,
has the form
\begin{equation}
\label{eq:free-hamil}
\hat{H} = - \frac{\hbar^{2}}{2M}\nabla^{2}_{\mathbf{R}} + \frac{B\hat{J}^{2}}{\hbar^{2}} \, ,
\end{equation}
with $M$ the total mass of the diatomic molecular ion. The derivatives
in the first term are with respect to the position of the center of
the mass of the ion, $\mathbf{R}$. The second term represents the
rotational energy of the ion, with $\hat{J}$ the total angular
momentum operator, the operator $\hat{J}^{2}$ having eigenvalues
$J(J+1)\hbar^{2}$, and the rotational constant is
\begin{equation}
  B = \frac{\hbar^{2}}{2I},
\end{equation}
where $I = m_{\mathrm{red}} r^{2}$ is the moment of inertia, with
$m_{\mathrm{red}}$ and $r$ the reduced mass and internuclear distance,
respectively.  The rigid rotor approximation we are using implies that
the latter is fixed, \ie, $r = \mathrm{const.}$

In addition to having a non-zero electric charge, we take the diatomic
molecular ion to be heteronuclear, which means that it possesses a
permanent dipole moment that will also interact with the electric field
of the Paul trap.  Adding the interaction of the ion with the trapping
field to Hamiltonian~(\ref{eq:free-hamil}), we get
\begin{equation}
\label{eq:whole-hamil}
\hat{H} = - \frac{\hbar^{2}}{2M}\nabla^{2}_{\mathbf{R}} +
\frac{B\hat{J}^{2}}{\hbar^{2}} + Ze\Phi(\mathbf{R},t) - \bm{\mu} \cdot
\mathbf{E}(\mathbf{R},t)\, , 
\end{equation}
in which $Ze$ is the total charge of the ion (with $e$ the elementary
charge), $\bm{\mu}$ is the permanent dipole moment, of magnitude
$\mu_{0}$, and $\mathbf{E} \equiv - \nabla \Phi$ is the trapping electric
field, at the position $\mathbf{R}$ of the ion.

\subsection{Classical Hamiltonian}
\label{subsec:CH}

Classically, we approximate the diatomic molecular ion inside a
trapping potential as a dipolar rigid rotor, consisting of two masses
$m_{-}$ and $m_{+}$ with partial charges $\delta_{-}$ and
$\delta_{+}$, respectively, kept spatially separated at a constant
distance.  A complete derivation of the kinetic and potential energy
terms is given in Appendix~\ref{app:model}, resulting in the
classical Hamiltonian
\begin{equation}
\label{eq:class-hamil}
H = K + V = 
\frac{1}{2} M \dot{\mathbf{R}}^2 + \frac{1}{2} I
  \dot{\bm{\omega}}^2 + Ze\Phi(\mathbf{R},t) - \bm{\mu} \cdot
  \mathbf{E}(\mathbf{R},t)\, ,
\end{equation}
in the point-dipole approximation, with $I$ the moment of inertia and
$\dot{\bm{\omega}}$ the angular velocity.  The corresponding classical
equation of motion for the center of mass is
\begin{equation}
  \label{eq:C_E_M}
  M\mathbf{\ddot{R}} + Ze\bm{\nabla}\Phi(\mathbf{R},t) -
  \bm{\nabla}\left [ \bm{\mu} \cdot \mathbf{E}(\mathbf{R},t) \right
  ]=0\, ,
\end{equation}
which differs from that of a classical atomic
ion\cite{Ghosh_book_1995,Major_book_2005,Hashemloo_IJMPC_2016} in an
additional term representing the coupling of the electric field with
the permanent dipole moment.  As shown in many previous works, the
equations of motion of an atomic ion inside a Paul trap obey the
Mathieu equation\cite{Ghosh_book_1995,Major_book_2005}
\begin{equation} 
\frac{\rm d^{2} \mathbf{R} }{\rm d \tau^{2} } + (a - 2q \cos 2\tau
)\mathbf{R} = 0\, . 
\label{eq:Math.eq}
\end{equation}
The Mathieu equation has stable solutions and hence bounded
trajectories depending on whether its parameters ($a$ and $q$) fall
within a certain range -- called stability regions -- in the $a-q$
plane or not.\cite{Wolf_NIST_2010} For an ion inside a Paul trap, $a$
and $q$ depend on the mass of the ion, the magnitude of the field, and
the parameters of the trap. For an ion in a linear Paul
trap, corresponding to the potential in Eqs.~(\ref{eq:tr_pot})
and~(\ref{eq:tr_pots}), setting $\tau = \Omega t/2$, we 
have
\begin{subequations}
\label{eq:a&q}
\begin{align} 
a_{x} = a_{y} &= -\frac{4Ze}{M\Omega^{2}}\frac{\kappa U_{0}}{z_{0}^{2}},\\
a_{z} &= \frac{8Ze}{M\Omega ^{2}}\frac{\kappa U_{0}}{z_{0}^{2}},\\
q_{x} = -q_{y} &= -\frac{2Ze}{M\Omega^{2}} \frac{V_{0}}{r_{0}^{2}},\\
q_{z} &= 0\, .
\end{align}
\end{subequations}
Even though the equation of motion for a diatomic molecular ion,
Eq.~(\ref{eq:C_E_M}), does not reduce to a Mathieu equation, taking the
term $\bm{\nabla}\left [ \bm{\mu} \cdot \mathbf{E}(\mathbf{R},t)
\right]$ as a small perturbation (we will investigate this below), one
can still use the stability criteria of the Mathieu equation as a
guideline for the stability of the trap for a molecular ion.

\section{Numerical methods}
\label{sec:num}

\subsection{Classical approach}
\label{subsec:class_num}

\begin{figure}
\centering
\includegraphics[scale=0.5]{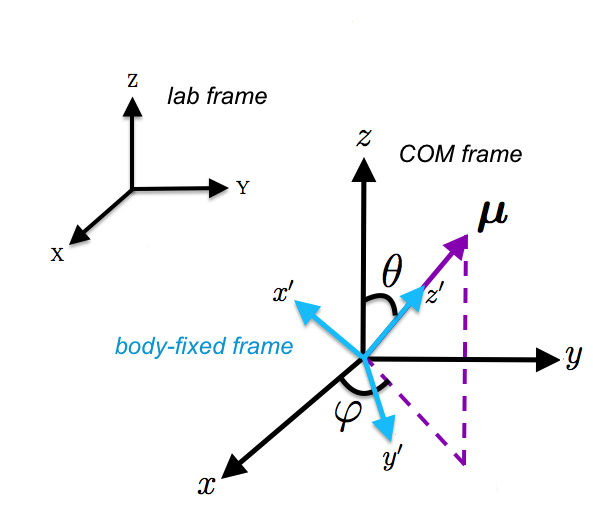}
\caption{(Color online) The orientation of the permanent dipole moment
  $\bm{\mu}$ of a diatomic molecule in space with respect to the
  center-of-mass (COM) frame $(x,y,z)$, which is indicated by angles $\varphi$
  and $\theta$. The body-fixed frame $({x}', {y}', {z}')$ shows the
  principal axes (or symmetry axes) of this molecule. The orientation
  of the dipole moment is fixed along ${z}'$.  The COM frame is
  aligned with the laboratory frame $(X,Y,Z)$, but has its origin at
  the center-of-mass of the molecule.}
\label{fig:Sph_pol}
\end{figure}

The classical representation of the rigid rotor,
Hamiltonian~(\ref{eq:class-hamil}), has five degrees of freedom:  the
three Cartesian coordinates of the center of mass, along with two
angles positioning the internuclear axis of the diatomic ion in
space, see Fig.~\ref{fig:Sph_pol}.
We consider three different coordinate systems: the laboratory-fixed
system $(X,Y,Z)$, the COM (center-of-mass) system $(x,y,z)$, which is
taken to be parallel to the laboratory-fixed frame, but centered on
the center of mass of the molecular ion (see also
Appendix~\ref{app:model}), and the body-fixed system
$({x}',{y}',{z}')$, with its origin at the center of mass, but with
axis ${z}'$ always co-linear with the internuclear axis (and hence
with the dipole moment $\bm{\mu}$).

Expanding the dipole moment vector into its components in the COM
frame $(x,y,z)$, we get
\begin{equation}
\label{eq:mu}
\bm{\mu} = \mu_{0} [\sin\theta \cos\varphi \mathbf{\hat{x}} +
\sin\theta \sin\varphi \mathbf{\hat{y}} + \cos\theta \mathbf{\hat{z}} ]\, ,
\end{equation}
in which $\mu_{0}$ is the magnitude of the dipole moment and
$\mathbf{\hat{x}}$, $\mathbf{\hat{y}}$, and $\mathbf{\hat{z}}$ are
unit vectors along the axes and $\theta$ and $\varphi$ are the polar
and azimuthal angles, respectively (see Fig.~\ref{fig:Sph_pol}). We
can also expand the trapping electric field at the position
of the ion into its component as
\begin{equation}
  \label{eq:electric-field}
  \mathbf{E}(\mathbf{R},t) = -\nabla \Phi(\mathbf{R},t) = \left[\frac{\kappa
      U_{0}}{z_{0}^{2}} - \frac{V_{0}}{r^{2}_{0}} \cos\Omega t
  \right]X \mathbf{\hat{x}} + \left[\frac{\kappa U_{0}}{z_{0}^{2}} +
    \frac{V_{0}}{r^{2}_{0}} 
    \cos\Omega t \right]Y \mathbf{\hat{y}} - \frac{2\kappa U_{0}}{z^{2}_{0}}Z
  \mathbf{\hat{z}} \, , 
\end{equation}
where we have used the potential defined by Eqs.~(\ref{eq:tr_pot}) and
(\ref{eq:tr_pots}).  The coupling term of Eq.~(\ref{eq:class-hamil})
then has the form
\begin{align}
\label{eq:int_term}
\bm{\mu} \cdot \mathbf{E} &= \left[\frac{\kappa
    U_{0}\mu_{0}\sin\theta\cos\varphi}{z_{0}^{2}} -
  \frac{V_{0}\mu_{0}\sin\theta\cos\varphi}{r^{2}_{0}} \cos\Omega t
\right]X \nonumber \\
&\quad + \left[\frac{\kappa
    U_{0}\mu_{0}\sin\theta\sin\varphi}{z_{0}^{2}}
  +\frac{V_{0}\mu_{0}\sin\theta\sin\varphi}{r^{2}_{0}} \cos\Omega t
\right]Y \\
&\quad - \left[\frac{2\kappa U_{0}\mu_{0}}{z^{2}_{0}}\cos\theta
\right] Z\, .\nonumber
\end{align}
It is clearly evident from this equation that the interaction between
the dipole moment and the electric field has not only a spatial
dependence, but also is dependent on the orientation of the ion inside
the trap.

A direct numerical simulation of the equations of motion corresponding
to the Hamiltonian (\ref{eq:class-hamil}) is impractical, as these
equations contain a term in $1/\sin^2\theta$, and thus are singular at
$\theta=0$. In order to avoid this problem, we use the ``body vector
method,''\cite{evans1977representatation} where it is assumed that the
diatomic molecular ion is actually a rigid rotor (di-axes body) with a
principal inertia tensor of the form\cite{evans1977representatation}
\begin{align}
\label{eq:inertia_tens}
I = \begin{pmatrix}
I_{x'} &0  &0 \\ 
 0 & I_{y'}  &0 \\ 
 0 &0 &0 
\end{pmatrix}\,
\end{align}
in which $I_{x'} = I_{y'} = I \equiv m_{\mathrm{red}} r^{2} $
($m_{\mathrm{red}}$ is the reduced mass and $r$ is the internuclear
distance) and $I_{z'} = 0$ (\ie, there is no rotation around the
internuclear axis), all expressed in the body-fixed coordinate system
$(x',y',z')$.  Following the body vector method, we introduce unit
vectors along these axes, namely $\mathbf{\hat{x}}'$,
$\mathbf{\hat{y}}'$, $\mathbf{\hat{z}}'$.  These unit vectors have components
with respect to the COM-frame
\begin{equation}
  \label{eq:unit_vec_com}
  \mathbf{\hat{u}}' = ({u}'_{x}, {u}'_{y}, {u}'_{z})\, .
\end{equation}
where $\mathbf{\hat{u}}' \in \left\{ \mathbf{\hat{x}}', \mathbf{\hat{y}}',
  \mathbf{\hat{z}}' \right\}$. 
  
Setting the internuclear
axis (or $\bm{\mu}$) along the $z'$ axis, the unit vector
$\mathbf{\hat{z}}'$ simply gives the direction of the dipole moment in
space.\cite{evans1977representatation} These vector are similar to
the variables described by Cheung\cite{Cheung_1, Cheung_2} to avoid
singularities in the equations of motion for diatomic molecules. As a
result, we can write the dipole moment vector as
\begin{equation}
  \label{eq:mu_vec}
  \bm{\mu} = \mu_{0}\mathbf{\hat{z}}'\, ,
\end{equation}
in which the components of $\mathbf{\hat{z}}'$ can be written in terms
of $\theta$ and $\varphi$
\begin{align}
\label{eq:unit_vec_z}
{z}'_x &= \sin \theta \cos \varphi\, , \nonumber\\
{z}'_y &= \sin \theta \sin \varphi\, ,\\
{z}'_z &= \cos \theta\, , \nonumber
\end{align}
in accord with Eq.~(\ref{eq:unit_vec_com}) and Fig.~\ref{fig:Sph_pol}.

We can now simply rewrite Eq.~(\ref{eq:int_term}) using the new
representation of the dipole moment, with the interaction term in the
Hamiltonian not being explicitly dependent on the angles $(\theta,
\varphi)$. Instead, the change in the direction of the dipole moment
is equivalent to the temporal change of the direction of the principal
axes according to
\begin{equation}
\label{eq:time_der_uni_vec}
\hat{\dot{\mathbf{u}}}' = \bm{\omega}_{P} \times \mathbf{\hat{u}}'\, ,
\end{equation}
where $\bm{\omega}_{P}$ is the principal angular velocity of the
dipole, precessing around its principal rotational axes. To write the
equations of motion, we need to calculate the torque in the lab-fixed
frame, $\mathbf{T}_{L} = \bm{\mu} \times \mathbf{E} $, which is
imposed on the dipole moment by the external electric
field.\cite{evans1977singularity} To calculate the time evolution of
the dipole moment, Eq.~(\ref{eq:time_der_uni_vec}), we need the
principal torque $\mathbf{T}_{P}$, which is obtained from the
relation\cite{evans1977representatation} 
\begin{equation}
\label{eq:transition_lab_mol}
\mathbf{T}_{L} = \mathbf{A}^{T} \mathbf{T}_{P}\, , 
\end{equation}
where $\mathbf{A}^{T}$ is the transpose of the rotation
matrix\cite{goldstein2002classical} whose columns are the principal
axes unit vectors\cite{evans1977representatation}
\begin{equation}
\label{eq:trans_matrix}
\mathbf{A}^{T} = (\mathbf{\hat{x}}', \mathbf{\hat{y}}',
\mathbf{\hat{z}}') 
\equiv \begin{pmatrix} 
  x'_{x} & y'_{x} & z'_{x}\\ 
  x'_{y} & y'_{y} & z'_{y}\\ 
  x'_{z} & y'_{z} & z'_{z}
\end{pmatrix}\, .
\end{equation}
As there is no rotation along the internuclear axis, $\omega_{P z'} =
0$ and there is no torque in the $z_P$ direction [see
Eq.~(\ref{eq:mu_vec})], $T_{P z'} = 0$, the equations of motion for
rotation are written as
\begin{align}
  \label{eq:torque}
  I_{x'}\dot{\omega}_{Px'} &= T_{Px'} +
  \omega_{Py'}\omega_{Pz'}(I_{y'}-I_{z'}) = T_{Px'}\nonumber \\ 
  I_{y'}\dot{\omega}_{Py'} &= T_{Py'} +
  \omega_{Pz'}\omega_{Px'}(I_{z'}-I_{x'}) = T_{Py'}\\ 
  I_{z'}\dot{\omega}_{Pz'} &=0\, .\nonumber
\end{align}

Finally, the equations which describe the dynamical
behavior of the ion inside the linear Paul trap, classically, are
\begin{subequations}
\label{eq:equations_of_motion}
\begin{align}
\frac{\rmd \mathbf{R}}{\rmd t} &= \frac{\mathbf{P}}{M}\,
, \label{eq:COM_classicalR}\\ 
\frac{\rmd P_{X}}{\rmd t} &= e \left [ b_{1} - b_{2} \cos \Omega t \right ]X + \mu_{0}\left [ b_{1} - b_{2} \cos \Omega t \right ]z'_{x}\, , \\
\frac{\rmd P_{Y}}{\rmd t} &= e \left [ b_{1} + b_{2} \cos \Omega t \right ]Y + \mu_{0}\left [ b_{1} + b_{2} \cos \Omega t \right ]z'_{y}\, , \\
\frac{\rmd P_{Z}}{\rmd t} &= - 2 e b_{1} Z -2 \mu_{0} b_{1} z'_{z}\,
, \label{eq:COM_classicalP} \\
\frac{\rmd {\mathbf{\hat{u}}}'}{\rmd t} &= \bm{\omega}_{P} \cdot
{\mathbf{\hat{u}}}'\, , \text{ where } \mathbf{\hat{u}}' \in \left\{
  \mathbf{\hat{x}}',\mathbf{\hat{z}}' \right\} \\
\intertext{(since only two coordinates are necessary to locate the
  dipole in space, we obtain $\mathbf{\hat{y}}'$ from $\mathbf{\hat{y}}' =
  \mathbf{\hat{z}}' \times \mathbf{\hat{x}}'$),}
\frac{\rmd \bm{\omega}_{P}}{\rmd t} &= \frac{1}{I} \mathbf{T}_{P}\ =
\frac{1}{I} ({\mathbf{A}^{\mathrm{T}}})^{-1} \mathbf{T}_{L}\,
, 
\end{align}
\end{subequations}
with
\begin{subequations}
\begin{align}
T_{Lx} &= \mu_{0} \left [ - 2 b_{1} Z z'_{y} - (b_{1} + b_{2} \cos \Omega t) Y z'_{z} \right ]\, , \\
T_{Ly} &= \mu_{0} \left [ 2 b_{1} Z z'_{x} + (b_{1} - b_{2} \cos \Omega t) X z'_{z} \right ]\, , \\
T_{Lz} &= \mu_{0} \left [ (b_{1} + b_{2} \cos \Omega t) Y
  z'_{x} - (b_{1} - b_{2} \cos \Omega t) X z'_{y} \right ]\, ,
\end{align}
\end{subequations}
where we have defined 
\begin{equation}
  \label{eq:b1&b2}
  b_{1} \equiv \frac{\kappa U_{0}}{z_{0}^{2}}\, , \quad
  b_{2} \equiv \frac{V_{0}}{r_{0}^{2}}\, .
\end{equation}

\subsection{Semi-classical approach}
\label{subsec:semi_class_num}

A direct simulation using the full quantum
Hamiltonian~(\ref{eq:whole-hamil}) would be quite numerically
intensive, owing to the five degrees of freedom, the large amplitude
of the motion in the trap, and the impossibility of separating the
problem along Cartesian coordinates (in contrast to the case for atomic
ions\cite{Hashemloo_IJMPC_2016}) due to the coupling of the rotation
with the trapping potential. Using the fact that the expectation value
for the center-of-mass motion follows the classical equations of
motion,\cite{Combescure_AIHPA_1986,Brown_PRL_1991,Glauber_1992,Hashemloo_IJMPC_2016}
we instead build a semi-classical model, where the center-of-mass
motion is treated classically, but the fully-quantum rotational wave
function is calculated, according to the Schr\"{o}dinger equation
\begin{equation}
\label{eq:Rot_Schrodinger}
\rmi \hbar \frac{\rmd }{\rmd t}\Psi^{\mathrm{rot}}(\theta,\varphi;t) =
\left [ \frac{B}{\hbar} \hat{J}^2 -\bm{\mu} \cdot
  \mathbf{E}(\mathbf{R},t) \right ] 
\Psi^{\mathrm{rot}}(\theta,\varphi;t)\, , 
\end{equation}
where $\Psi^{\mathrm{rot}}(\theta,\varphi;t)$ is the rotational wave
function of the molecular ion and $\mathbf{E}(\mathbf{R},t)$ is the
trapping electric field at the position of the ion,
$\mathbf{R}$, obtained classically, see
Eqs.~(\ref{eq:eq_mot_semi_class}) below.

The wave function
can be expanded on the basis of
field-free rotor states (spherical harmonics) $Y_{J,M} (\theta,\varphi)$ as
\begin{equation}
  \label{eq:rot_function}
  \Psi^{\mathrm{rot}}(\theta,\varphi;t) = \sum_{J,M} D_{J,M}(t) Y_{J,M}(\theta,\varphi)\, ,
\end{equation} 
where the expansion coefficients $D_{J,M}$ are time dependent complex
values that determine the rotational state of the ion.

Solving the time-dependent Schr\"{o}dinger
equation~(\ref{eq:Rot_Schrodinger}) using the expression of the
rotational wave function in Eq.~(\ref{eq:rot_function}),
\begin{equation}
  \label{eq:eq_mot_coeff_1}
  \rmi \hbar\sum_{J,M}\dot{D}_{J,M}(t) = \sum_{J,M}D_{J,M}(t)BJ(J+1)
  -\sum_{J,M} \sum_{{J}',{M}'} D_{{J}',{M}'}(t) \braket{J,M |
  \bm{\mu}\cdot\mathbf{E} | {J}',{M}'} 
  \, ,
\end{equation}
results in a set of coupled equations for the time evolution of the
coefficients, 
\begin{align}
  \label{eq:eq_mot_coeff_2}
  \dot{D}_{J,M}(t) = - \frac{\rmi}{\hbar}D_{J,M}(t) BJ(J+1)
  +\frac{\rmi}{\hbar}\sum_{{J}',{M}'} D_{{J}',{M}'}(t) \braket{
    J,M|\bm{\mu}\cdot\mathbf{E}|{J}',{M}'} \, , 
\end{align}
in which the kets $\ket{J,M}$ correspond to the spherical harmonics
$Y_{J,M} (\theta,\varphi)$. Using Eq.~(\ref{eq:int_term}) for the
interaction of the dipole moment with the electric field and
substituting trigonometric functions with their equivalent spherical
harmonics\cite{varshalovich1988quantum} gives
\begin{equation}
\label{eq:interaction_spher}
\bm{\mu}\cdot\mathbf{E} = \sqrt{\frac{2\pi}{3}}\mu_{0} \left[ (E_{x} + \rmi
E_{y})Y_{1,-1} + \sqrt{2}E_{z}Y_{1,0} + (-E_{x} + \rmi E_{y})Y_{1,1} \right]
\end{equation}
and hence
\begin{align}
\label{eq:expec_inter}
\braket{J,M | \bm{\mu}\cdot\mathbf{E} | {J}',{M}'} &=
\sqrt{\frac{2\pi}{3}}\mu_{0} \left[ \left( E_{x} + \rmi E_{y} \right) \braket{J,M | Y_{1,-1} |
  {J}',{M}'} + \sqrt{2}E_{z} \braket{J,M | Y_{1,0} | {J}',{M}'}
\right. \nonumber \\ 
&\quad \left. + \left( -E_{x} + \rmi E_{y} \right) \braket{J,M |
    Y_{1,1} | {J}',{M}'} 
\right] \, ,  
\end{align}
Using Clebsch-Gordan coefficients and their related 3j-symbols, we have\cite{varshalovich1988quantum}
\begin{align}
\label{eq:Cleb_Gor}
\braket{J,M | Y_{1,n} | {J}',{M}'} &=
\int_{0}^{\pi}\int_{0}^{2\pi}Y^*_{J,M}(\theta,\varphi)Y_{1,n}(\theta,\varphi)Y_{{J}',{M}'}(\theta,\varphi)\sin
\theta \rmd \theta \rmd \varphi \nonumber \\ &=
(-1)^{n}\sqrt{\left [\frac{(2J+1)(3)(2{J}'+1)}{4\pi}\right
  ]}\begin{pmatrix} 
J & 1 & {J}' \\ 
0 & 0 & 0
\end{pmatrix}\begin{pmatrix}
J & 1 & {J}' \\ 
-M & n & {M}'
\end{pmatrix} \nonumber \\ &=\sqrt{\left [\frac{(2{J}'+1)(3)}{4\pi(2J+1)} \right ]}C^{J0}_{{J}'010}C^{JM}_{{J}'{M}'1n}  \, , 
\end{align}
with $n = -1,0,1$. The selection rules for non-zero values of the the
Clebsch-Gordan coefficients in Eq.~(\ref{eq:Cleb_Gor}) imply the
relations
\begin{subequations}
\begin{align}
\label{eq:selection_rule}
\Delta J &\equiv \left | J - {J}' \right | = 0, \pm 1
  \,\,\, (J = 0 \not \leftrightarrow 0) \\ \Delta M &\equiv
  \left | M -     {M}' \right | = 0, \pm 1\, .
\end{align}
\end{subequations}

The classical equations for the motion of the center of mass were
given in the previous section,
Eqs.~(\ref{eq:COM_classicalR})--(\ref{eq:COM_classicalP}), and they
involve the 
orientation of the dipole through the vector $\mathbf{\hat{z}}'$,
Eq.~(\ref{eq:unit_vec_z}).  As $\theta$ and $\varphi$ are not, in the
semi-classical approach, classical variables, we need to modify the
above treatment for the center-of-mass motion.  To take into account
the effect of rotation, we calculate the orientation of the dipole using
\begin{subequations}
  \label{eq:expect_sin_cos}
  \begin{align}
    {z}'_{x} \equiv \left \langle \sin \theta \cos \varphi \right
    \rangle(t) &= \sum_{J,M} \sum_{{J}',{M}'}
    D^{*}_{J,M}(t)D_{{J}',{M}'}(t) \braket{J,M | \sin \theta \cos \varphi
      | {J}',{M}'} \, , \\
    {z}'_{y} \equiv \left \langle \sin \theta \sin \varphi \right
    \rangle(t) &= \sum_{J,M} \sum_{{J}',{M}'}
    D^{*}_{J,M}(t)D_{{J}',{M}'}(t) \braket{ J,M | \sin \theta \sin
    \varphi | {J}',{M}'} \, , \\
    {z}'_{z} \equiv \left \langle \cos \theta \right \rangle(t) &=
    \sum_{J,M} \sum_{{J}',{M}'} D^{*}_{J,M}(t)D_{{J}',{M}'}(t)
    \braket{J,M | \cos \theta | {J}',{M}'} \, , 
\end{align}
\end{subequations}
where we have indicated explicitly that the expectation values are
time-dependent variables. Finally, the time evolution of a rigid-rotor
diatomic molecular ion trapped in a linear Paul trap, in the
semi-classical model, is obtained from the coupled differential
equations
\begin{subequations}
\label{eq:eq_mot_semi_class}
\begin{align}
\frac{\rmd \mathbf{R}}{\rmd t} &= \frac{\mathbf{P}}{M}\, , \\
\frac{\rmd P_{X}}{\rmd t} &= e \left [ b_{1} - b_{2} \cos \Omega t
\right ]X + \mu_{0}\left [ b_{1} - b_{2} \cos \Omega t \right ]\left
  \langle \sin \theta \cos \varphi \right \rangle (t)\, , \\ 
\frac{\rmd P_{Y}}{\rmd t} &= e \left [ b_{1} + b_{2} \cos \Omega t
\right] Y + \mu_{0}\left [ b_{1} + b_{2} \cos \Omega t \right ]\left
  \langle \sin \theta \sin \varphi \right \rangle (t)\, , \\ 
\frac{\rmd P_{Z}}{\rmd t} &= - 2 e b_{1} Z -2 \mu_{0} b_{1}\left \langle \cos \theta \right \rangle (t)\, , \\
\frac{\rmd  D_{J,M}}{\rmd t} &= - \frac{\rmi}{\hbar}D_{J,M}(t) BJ(J+1) 
+\frac{\rmi}{\hbar}\sum_{{J}',{M}'} D_{{J}',{M}'}(t) \braket{ J,M |
  \bm{\mu}\cdot\mathbf{E} | {J}',{M}'} \, ,
\end{align}
\end{subequations}
with $b_{1}$ and $b_{2}$ defined by Eq.~(\ref{eq:b1&b2}).  Note that
this is different from the result that would be obtained by a simple
classical approximation of the momentum term in
Hamiltonian~(\ref{eq:whole-hamil}), as this would omit the effect of
the coupling term $\bm{\mu} \cdot \mathbf{E}$ on the center-of-mass
motion.

\subsection{Numerical implementation}

To solve the equations of motion, both for the classical and the
semi-classical models, we use a 4th-order Runge-Kutta-Fehlberg
integrator.\cite{pressC92,gsl} The time step is taken to be equal to
$\Delta t = \SI{1e-14}{\second}$.  In the classical simulations, this
choice of time step ensures that the unit vector $\mathbf{\hat{u}}'$ remains
normalised at each time step.

\section{Results}
\label{sec:res}

In both classical and semi-classical simulations we have used
$\mathrm{MgH}^{+}$ ion as an example. The magnitude of the permanent
dipole moment of the ion, $\mu_{0}$, as well as its rotational
constant, $B$, the mass of each of the consisting atoms,
$m_{\mathrm{H}}$ and $m_{\mathrm{Mg}}$, and the internuclear distance
$r$ are all listed in Tab.~\ref{tab:ion_param}.\cite{Drewsen_1, Japan}
\begin{table}
  \caption{\label{tab:ion_param}Characteristic values for the MgH$^+$
    ion.\cite{Drewsen_1, Japan}}  
  \begin{center}
    \begin{tabular}{cc}
      \hline \hline
      Parameter & Value \\ \hline
      $m_{\mathrm{H}}$ & \SI{1.6737e-27}{\kilo\gram} \\
      $m_{\mathrm{Mg}}$ & \SI{4.0359e-26}{\kilo\gram} \\
      $\mu_{0}$ &  \SI{3.6}{Debye}\\
      $B$ &  \SI{6.4058}{\per\centi\meter} \\
      $r$ &  \SI{1.655}{{\AA}} \\
      \hline \hline
    \end{tabular}
  \end{center}
\end{table}
We use one set of trap parameters for all simulations, except for the
amplitudes of the static and radio-frequency electric potentials
$U_{0}$ and $V_{0}$, that may vary in different numerical
calculations. The value of the fixed parameters are given in
Tab.~\ref{tab:trap_param} and are based on typical experimental
realizations.\cite{Raizen_PRA_1992}
\begin{table}
  \caption{\label{tab:trap_param}Trap configuration parameters for a
    MgH$^+$ ion.}  
  \begin{center}
    \begin{tabular}{cc}
      \hline \hline
      Parameter & Value \\ \hline
      $r_{0}$ &  \SI{0.769e-3}{\meter} \\
      $z_{0}$ &  \SI{1.25e-3}{\meter} \\
      $\kappa$ &  $0.31$ \\
      $\Omega$ &   $2\pi \times \SI{8e6}{\per\second}$ \\
      \hline \hline
    \end{tabular}
  \end{center}
\end{table}
With these parameters, we can calculate the stability regions of the
Mathieu equation, using a point-charge model, without the interaction
of the dipole moment with the trap (as discussed in
Sec.~\ref{subsec:CH}).  

\begin{figure}
\centering
\includegraphics[width=0.6\textwidth]{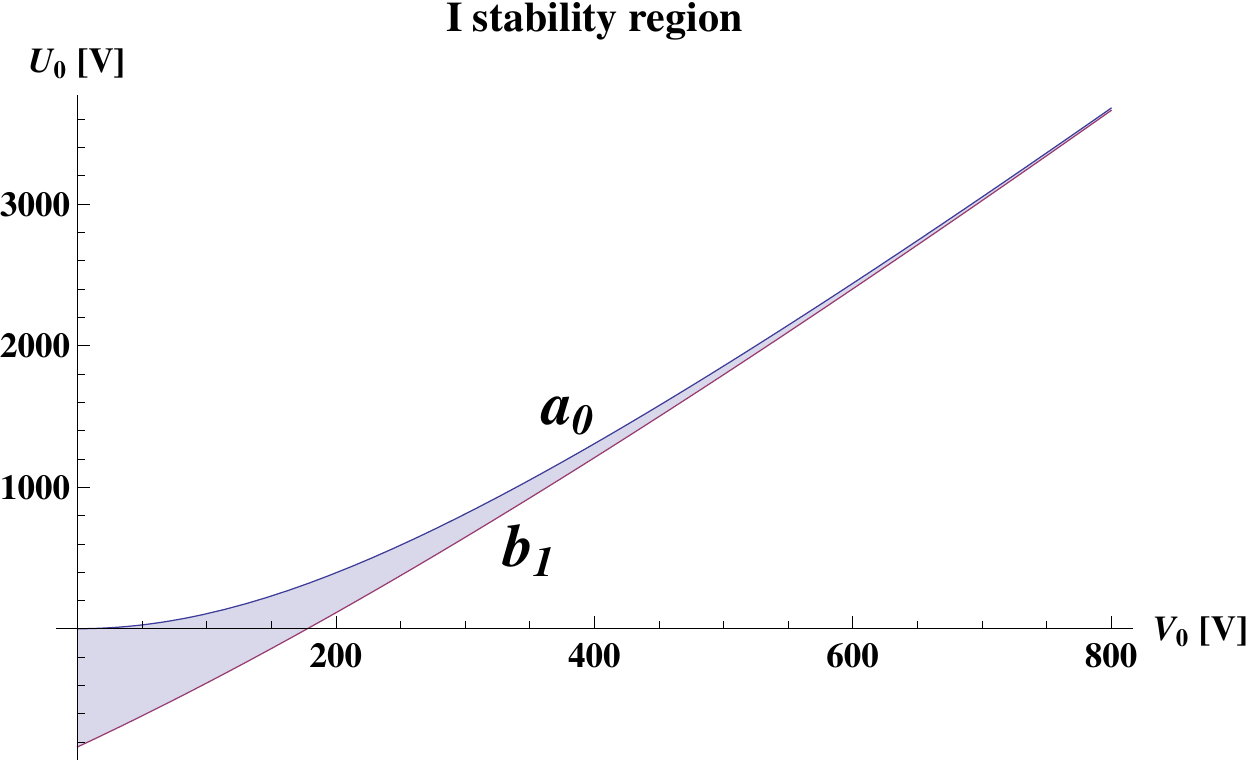}
\centering
\includegraphics[width=0.63\textwidth]{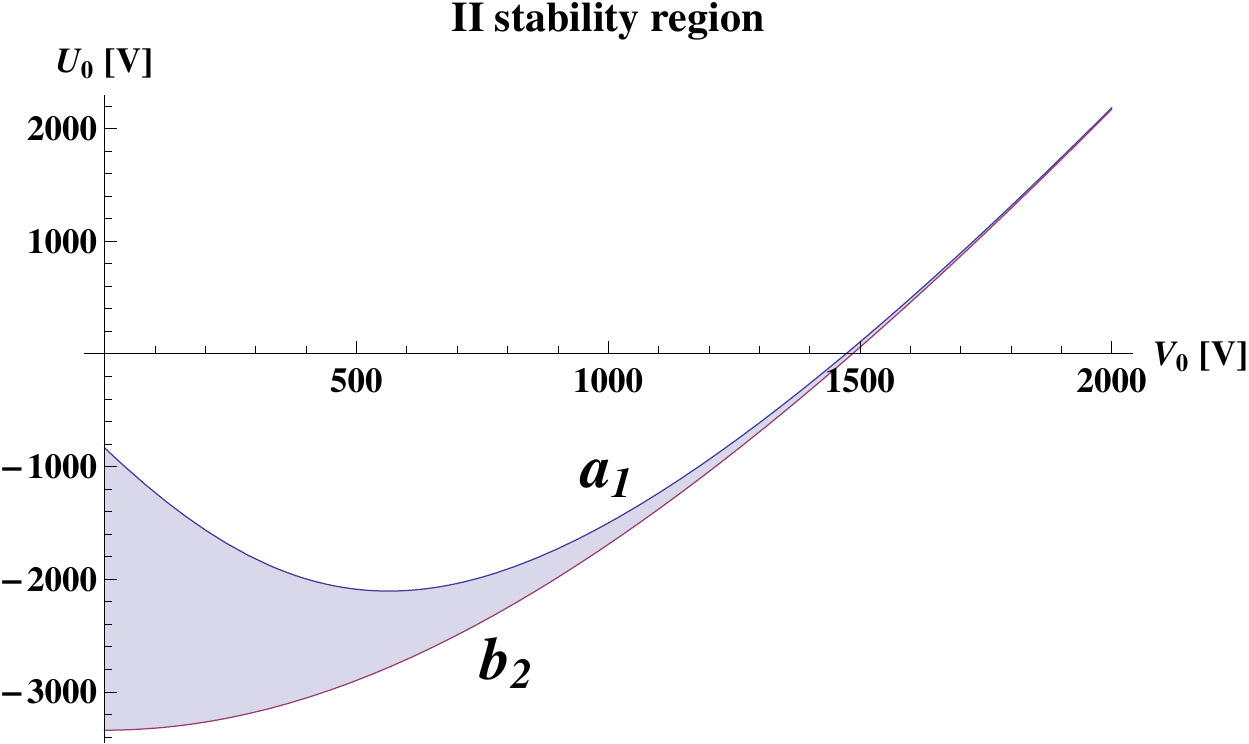}
\caption{\label{fig:stab}(Color online) I and II stability regions of
  the Mathieu equation, corresponding to a $\mathrm{MgH^{+}}$
  molecular ion, in a point-charge model. Shaded areas correspond to
  stable trajectories, with the borders delimited by the
  characteristic values of the Mathieu equation.\cite{Wolf_NIST_2010}}
\end{figure}

We ran our simulations for three different pairs of trapping
potentials $(U_{0}, V_{0})$ equal to $(\SI{680}{\volt},
\SI{1650}{\volt})$, $(\SI{2175}{\volt}, \SI{2000}{\volt})$ and
$(\SI{7080}{\volt}, \SI{3000}{\volt})$, resulting in values of $a$ and
$q$, Eqs.~(\ref{eq:a&q}), which are located inside the second
stability region of the Mathieu equation for an ion with no dipole
moment.  The I and II stability regions corresponding to a
$\mathrm{MgH^{+}}$ molecular ion in a linear Paul trap are shown in
Fig.~\ref{fig:stab}, as calculated from the characteristic values
$a_0$, $b_1$, etc., of the Mathieu equation.\cite{Wolf_NIST_2010} Note
that these stability regions hold along both $x$ and $y$, as they
depend on $a$ and $|q|$, while the trapping along $z$ is
unconditionally stable.

We chose our parameters from the second stability region since it
provides trapping potentials which are large enough to magnify the
effect of the coupling between the dipole moment and the trapping
field. However it needs to be mentioned that we also ran simulations
for much smaller trapping potentials [for example $(\SI{1820}{\volt},
\SI{500}{\volt})$], in the first stability region.  Although the
effect of the coupling between the dipole and potential field was not
as significant in the latter case, by changing the initial energy of
the ion, either by increasing its initial kinetic energy or
putting it off-center, we obtained results qualitatively similar to
the ones presented here.

In our simulations, the ion is located initially ($t=0$) at the center
of the trap.  However as we mentioned in the previous paragraph,
choosing an initial position other than the center of the trap gives
the same qualitative results.  We also assign an initial linear
momentum (in both methods) and an initial angular momentum to the ion
(in the classical method), approximating an initial temperature.  The
latter is chosen based on experimental works on trapping
$\mathrm{MgH}^{+}$ ions. For example, $\mathrm{MgH}^{+}$ ions have
been cooled translationally down to temperature of $\sim
\SI{10}{\milli\kelvin}$ by Doppler laser cooling where ions form a
Coulomb crystal, or temperatures of $\sim \SI{10}{\kelvin}$ are
obtained where they are cooled rotationally.\cite{Drewsen_2} In this
work, we have assigned the same initial temperature to both classical
rotational and translational motion of the ion, equivalent to
\SI{20}{\kelvin}.

\subsection{Classical trajectories of the center-of-mass motion of the ion}
\label{subsec:res_class}

In the classical simulations, in addition to assigning an initial
position and initial linear and angular momenta to the ion, we have
also assigned an initial orientation of the ion inside the linear Paul
trap with respect to the COM frame, whose $z$-axis is considered to be
parallel to the elongated part of the trapping device, in
correspondence with Eq.~(\ref{eq:unit_vec_z}) (see also
Fig.~\ref{fig:Sph_pol}). Consequently, we will consider three
different initial orientations of the dipole moment (internuclear
axis): i) along the $z$-axis ($\theta = 0^{\circ}$); ii) in the $x-y$
plane $(\theta = 90^{\circ}, \varphi = 45^{\circ})$; iii) along an
arbitrarily-chosen orientation in space $(\theta = 45^{\circ}, \varphi =
30^{\circ})$.  These three orientations are indicated by the angles
$\theta$ and $\varphi$ with respect to the COM frame, see
Fig.~\ref{fig:Sph_pol}.  Let us note that the simulation requires two
vectors, $\mathbf{\hat{x}}'$ and $\mathbf{\hat{z}}'$, see
Eqs.~(\ref{eq:equations_of_motion}), while $\theta$ and $\varphi$ only
fix $\mathbf{\hat{z}}'$.  We therefore take an arbitrary initial
orientation for $\mathbf{\hat{x}}'$, orthogonal to
$\mathbf{\hat{z}}'$.  We ran the classical simulations for a total
time of \SI{5}{\micro\second} and for three different pairs of
trapping potentials mentioned above.

\begin{figure}
\includegraphics[width=0.49\textwidth]{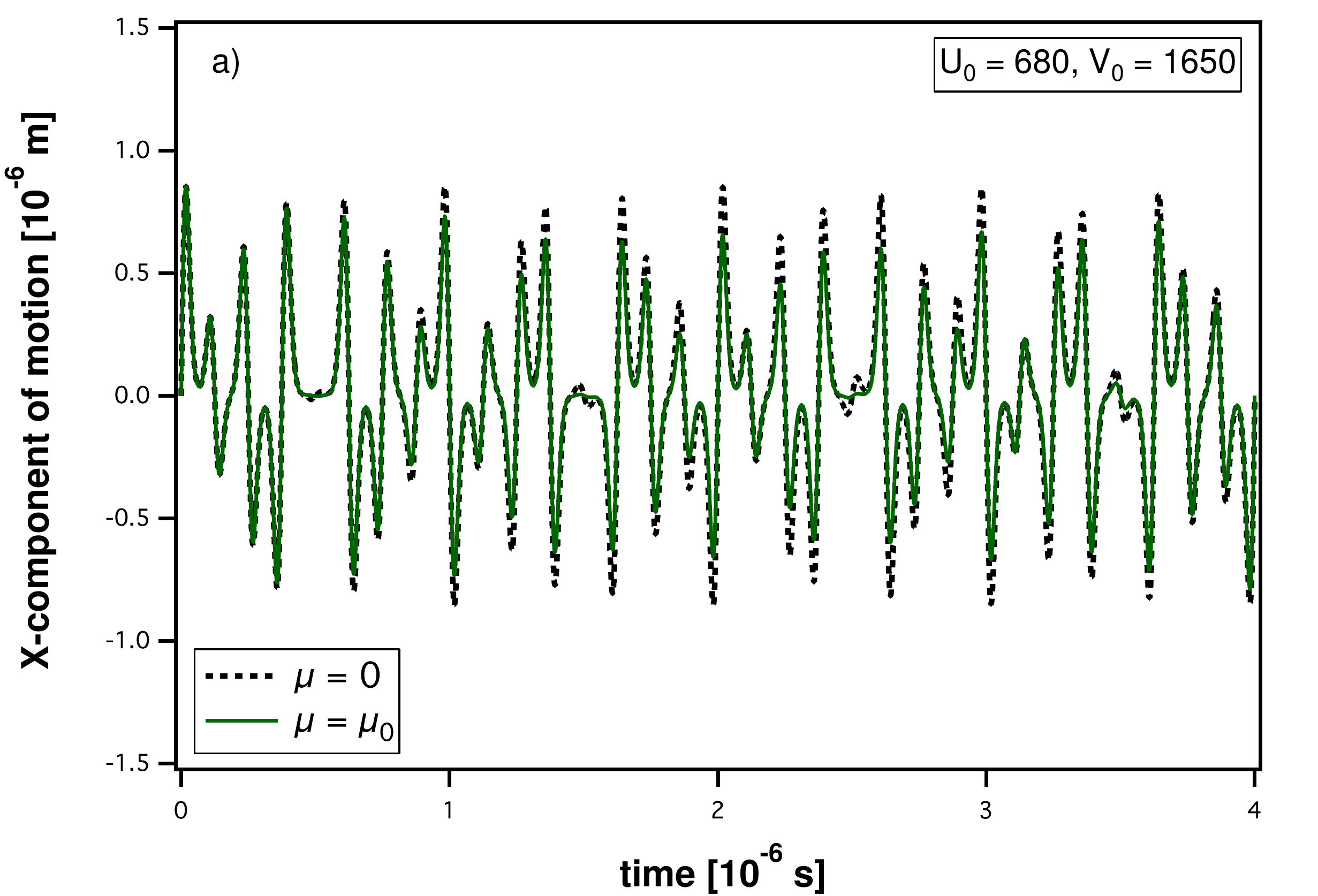}   
\includegraphics[width=0.49\textwidth]{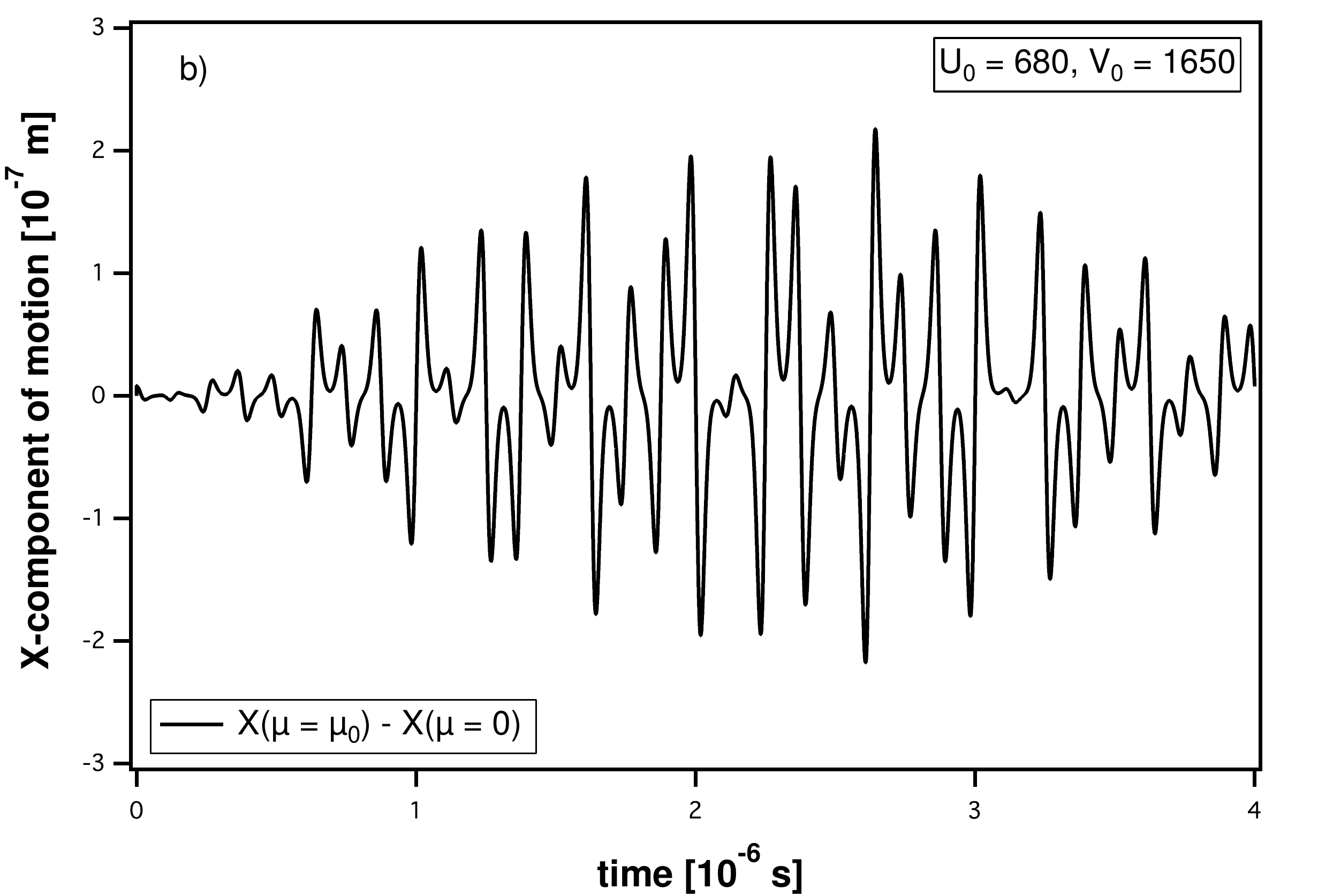} \\
\includegraphics[width=0.49\textwidth]{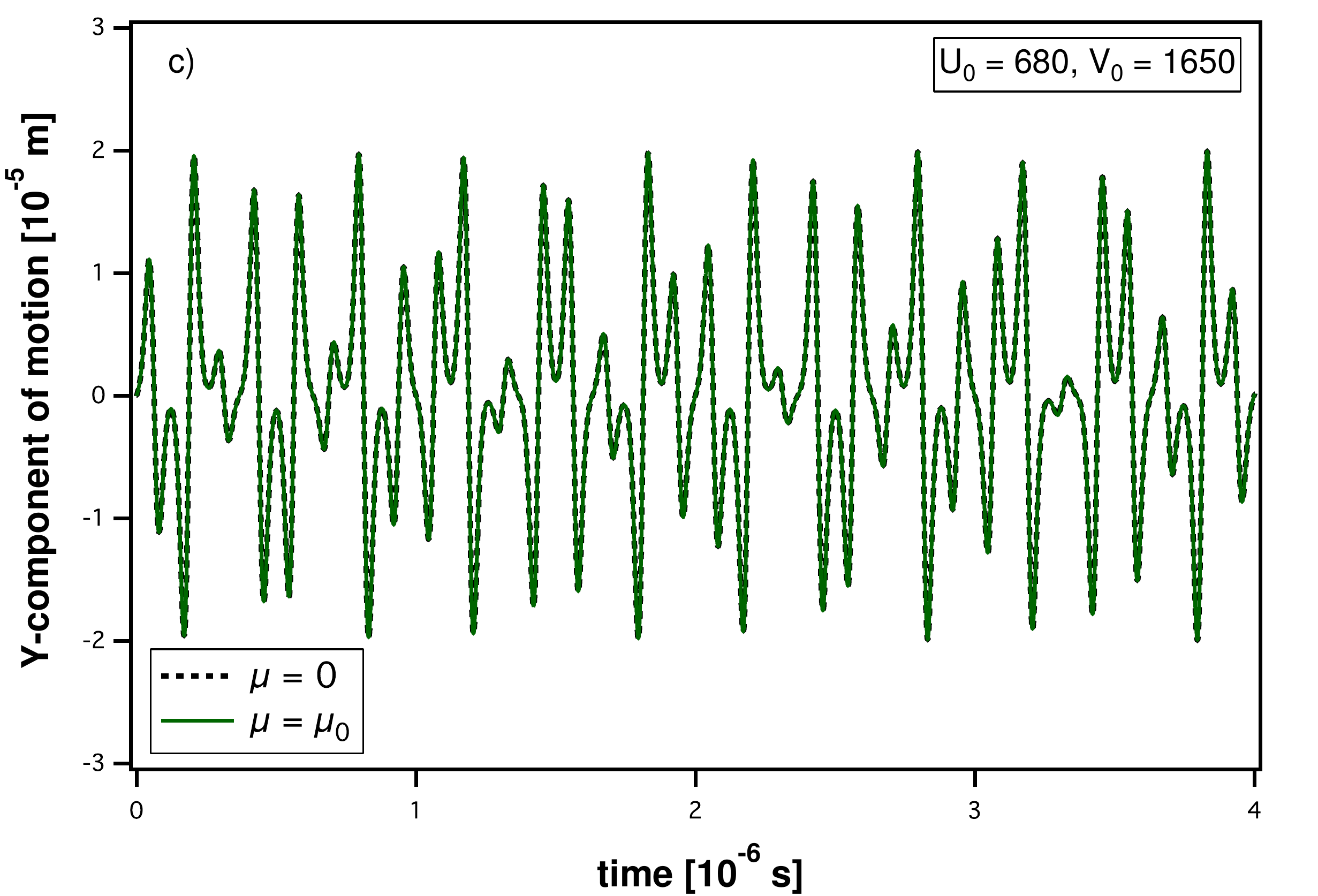}    
\includegraphics[width=0.49\textwidth]{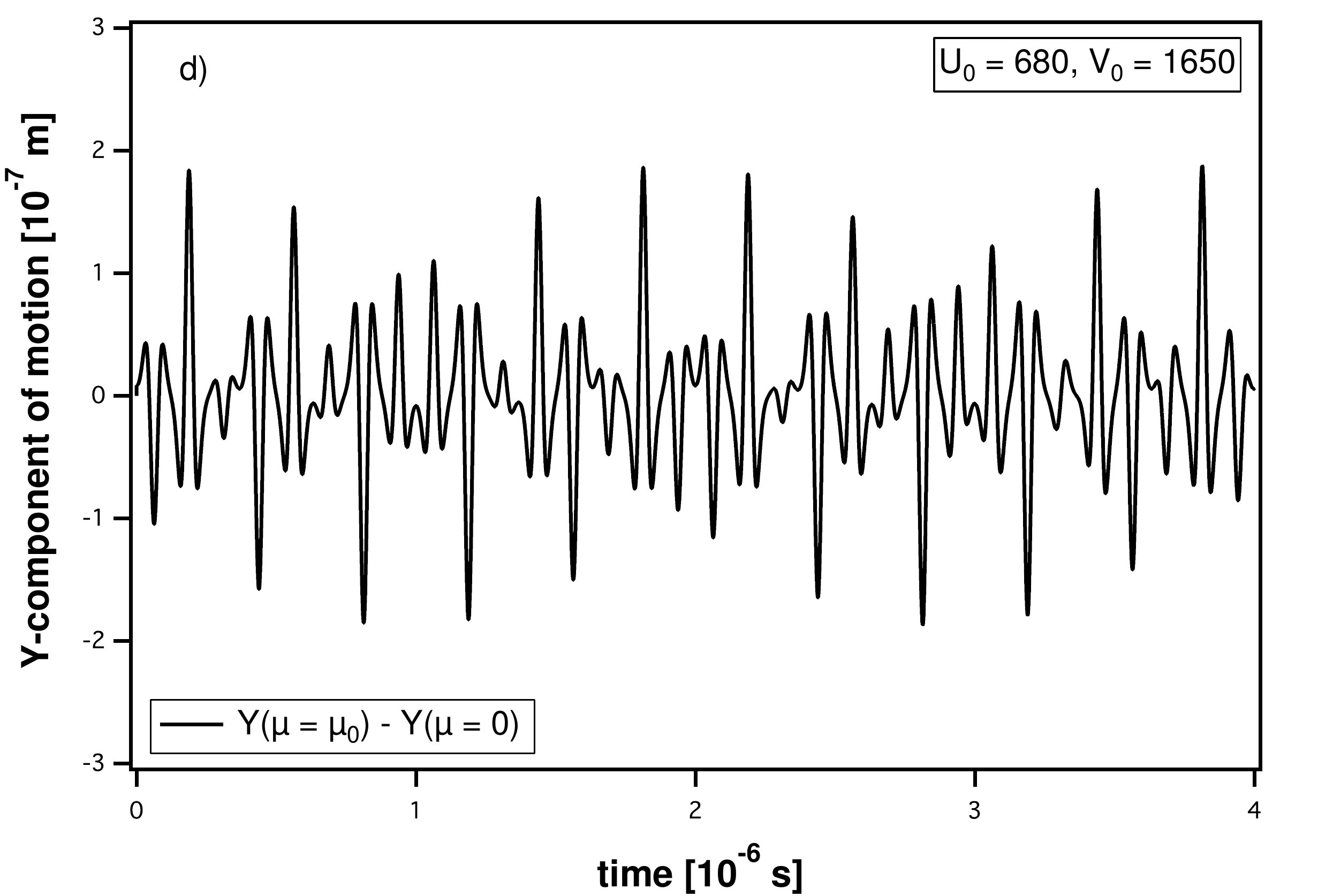} \\
\includegraphics[width=0.49\textwidth]{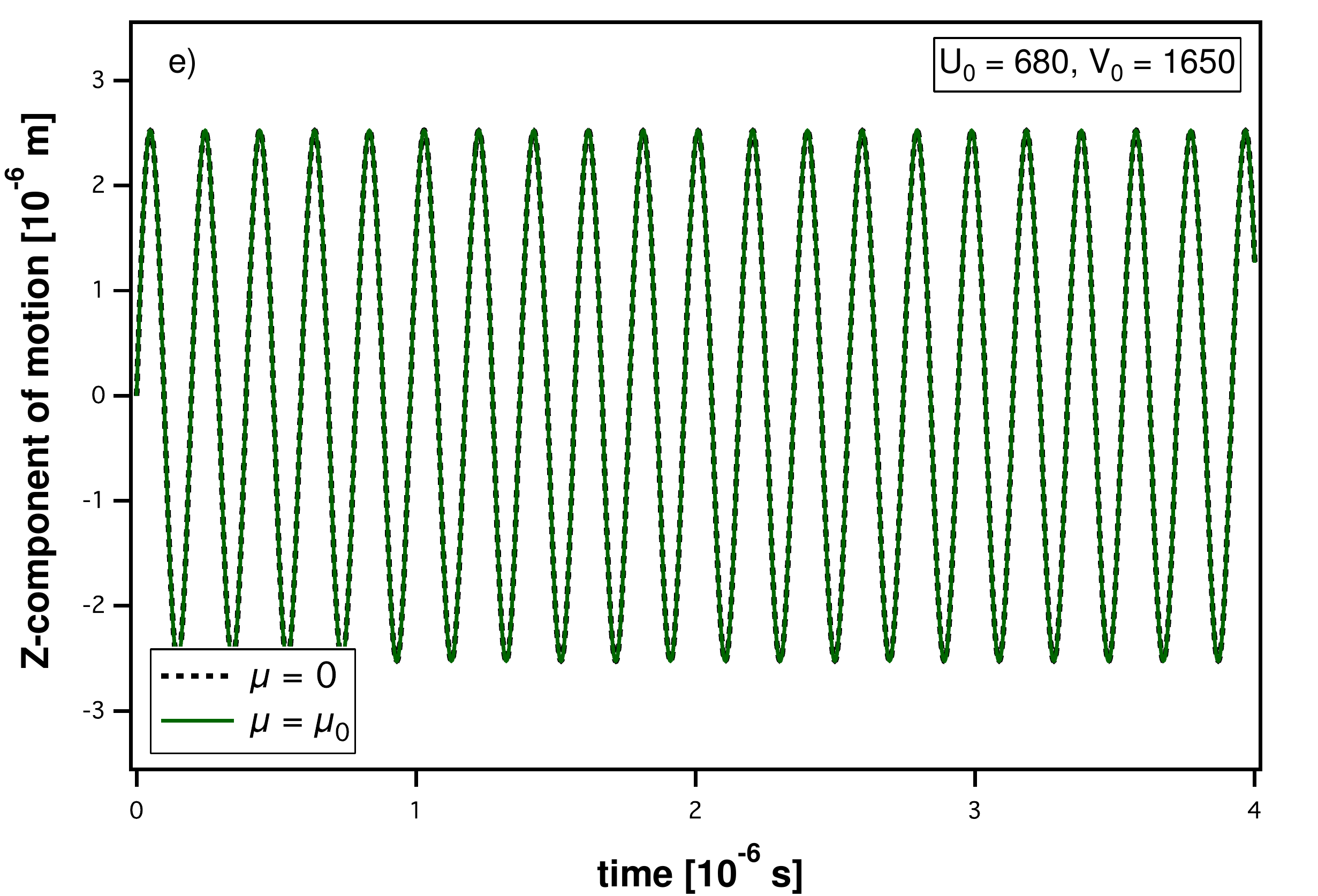}    
\includegraphics[width=0.49\textwidth]{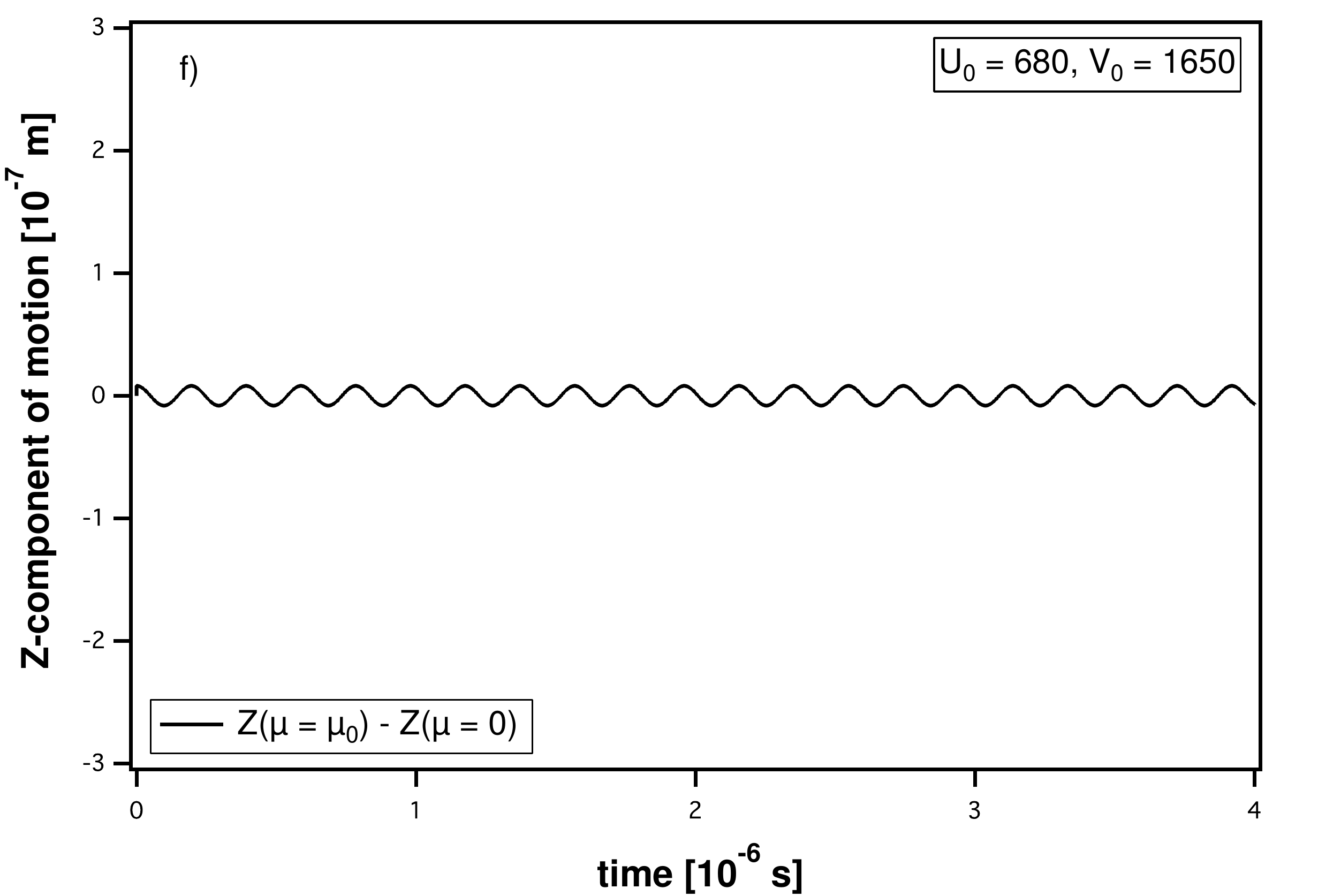}
\caption{(Color online) Time evolution of the center-of-mass motion of
  a $\mathrm{MgH^{+}}$ ion inside a linear Paul trap, with classical
  rotation: (a)--(b) $X$ component of the motion, (c)--(d)
  $Y$ component of the motion, and (e)--(f) $Z$ component of the
  motion. In panels (a), (c) and (e), the dotted (black) curves
  correspond to the trajectories of the ion when $\mu_{0} = 0$, while
  the solid (green) curves show the trajectories when $\mu_{0} \neq
  0$. Panels (b), (d), and (f) present the difference between the
  trajectories, with and without the dipole moment. The magnitudes of
  the trapping potentials are $U_{0} = \SI{680}{\volt}, V_{0} =
  \SI{1650}{\volt}$.}
\label{fig:class_first}
\end{figure}

The trajectories of the center-of-mass motion of the ion for two
cases, when there is an interaction between the dipole moment and the
electric field ($\mu \neq 0$) and when this interaction is ignored
($\mu = 0$), are shown for all three components of the motion in
Fig.~\ref{fig:class_first}. For these simulations, $(U_{0}, V_{0})$
are equal to $(\SI{680}{\volt}, \SI{1650}{\volt})$ and the diatomic
ion is initially oriented in space with spherical angles $\theta =
45^{\circ}$ and $\varphi = 30^{\circ}$.
As expected from Eqs.~(\ref{eq:equations_of_motion}), one sees that in
the presence of a permanent dipole moment, the trajectory of the
center of mass does not correspond to the Mathieu equation, which is
obtained for $\mu = 0$.  For such a strength of the trapping field and
initial orientation, the deviation is significant enough to be
observable in the $x$ direction.  However, in the $y$ and $z$
directions the difference between these trajectories is very small
compared to the actual amplitudes of the motion.  In order to magnify
the deviation in the trajectories, especially in the $y$
and $z$ directions, we have plotted the difference between the
trajectories with and without the dipole moment in
Figs.~\ref{fig:class_first}(b), (d), and (f).  In the $z$ direction,
this difference is small enough to be assumed equal to zero. This
arises from the fact that the equation of motion in the $z$ direction,
Eq.~(\ref{eq:equations_of_motion}d), lacks the term containing the
radio-frequency field, resulting only in a fast-oscillating term (due
to the time evolution of $z_z'$) that averages out to zero.  In all
cases, the presence of the coupling between the dipole moment and the
trapping field does not affect the stability of ion's trajectory in
the trap, which remains bounded and even displays the same oscillation
period as when there is no coupling.

\begin{figure}
\centering
\includegraphics[scale=0.33]{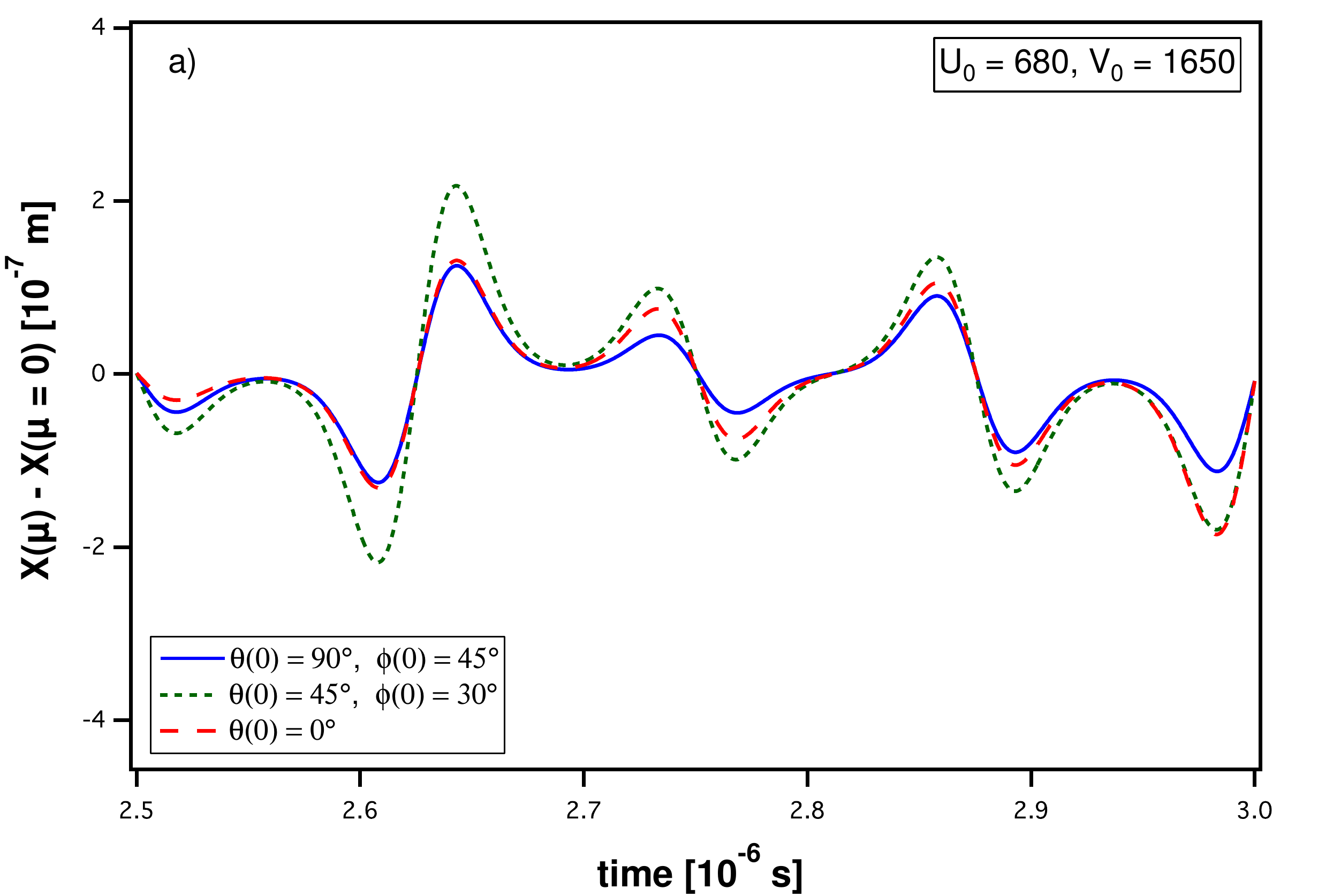}
\centering
\includegraphics[scale=0.33]{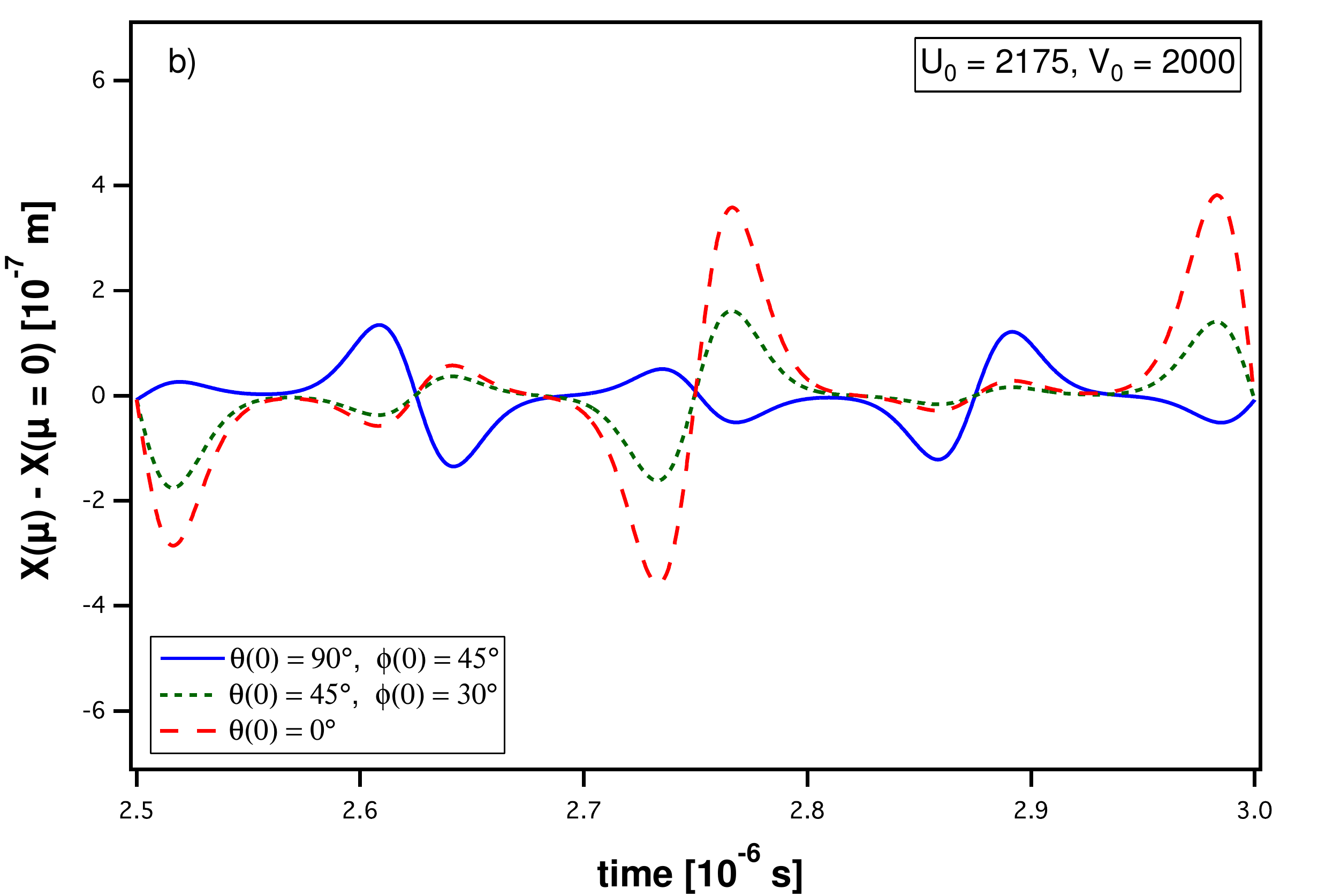}
\centering
\includegraphics[scale=0.33]{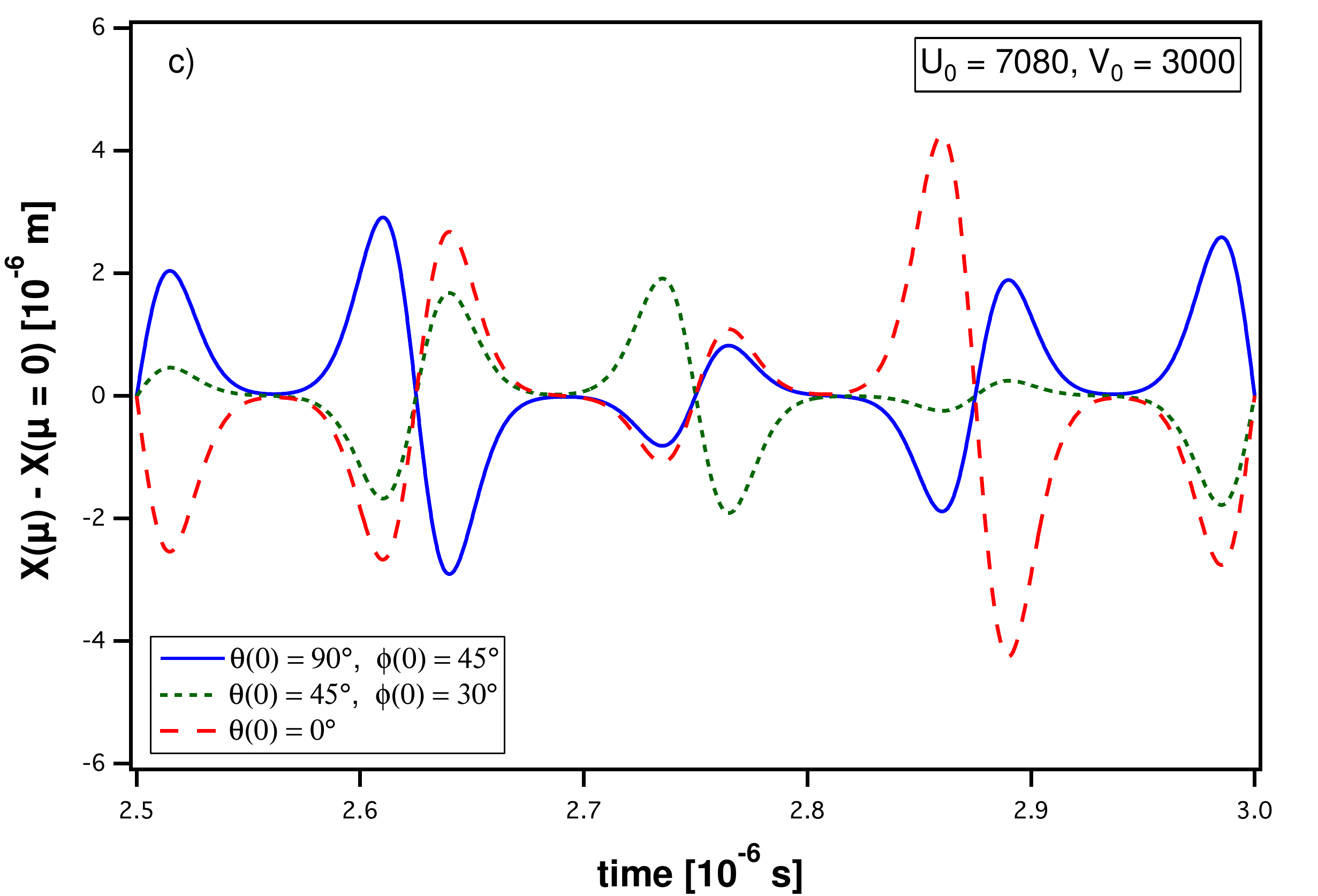}
\caption{(Color online) Time evolution of the $X$ component of the
  center-of-mass motion, with classical rotation, for different initial
  angles and different trap potentials: (a) $U_{0} = \SI{680}{\volt}$,
  $V_{0} = \SI{1650}{\volt}$; (b) $U_{0} = \SI{2175}{\volt}$, $V_{0} =
  \SI{2000}{\volt}$; (c) $U_{0} = \SI{7080}{\volt}$, $V_{0} =
  \SI{3000}{\volt}$. The curves present the difference between the
  trajectories of motion for the two cases of $\mu = 0$ and $\mu =
  \mu_{0}$.}
\label{fig:class_second}
\end{figure}

\begin{figure}
\centering
\includegraphics[scale=0.33]{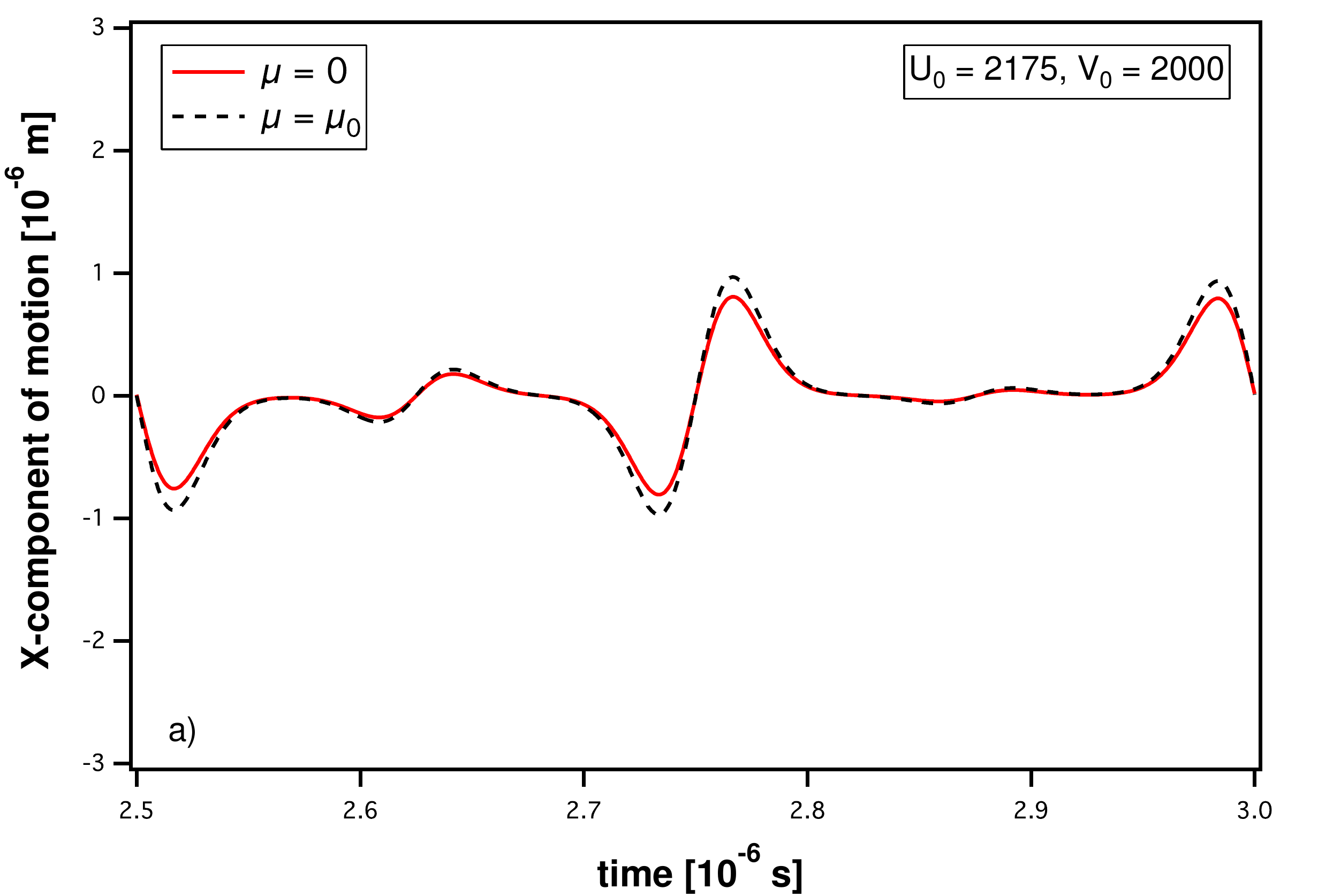}
\centering
\includegraphics[scale=0.33]{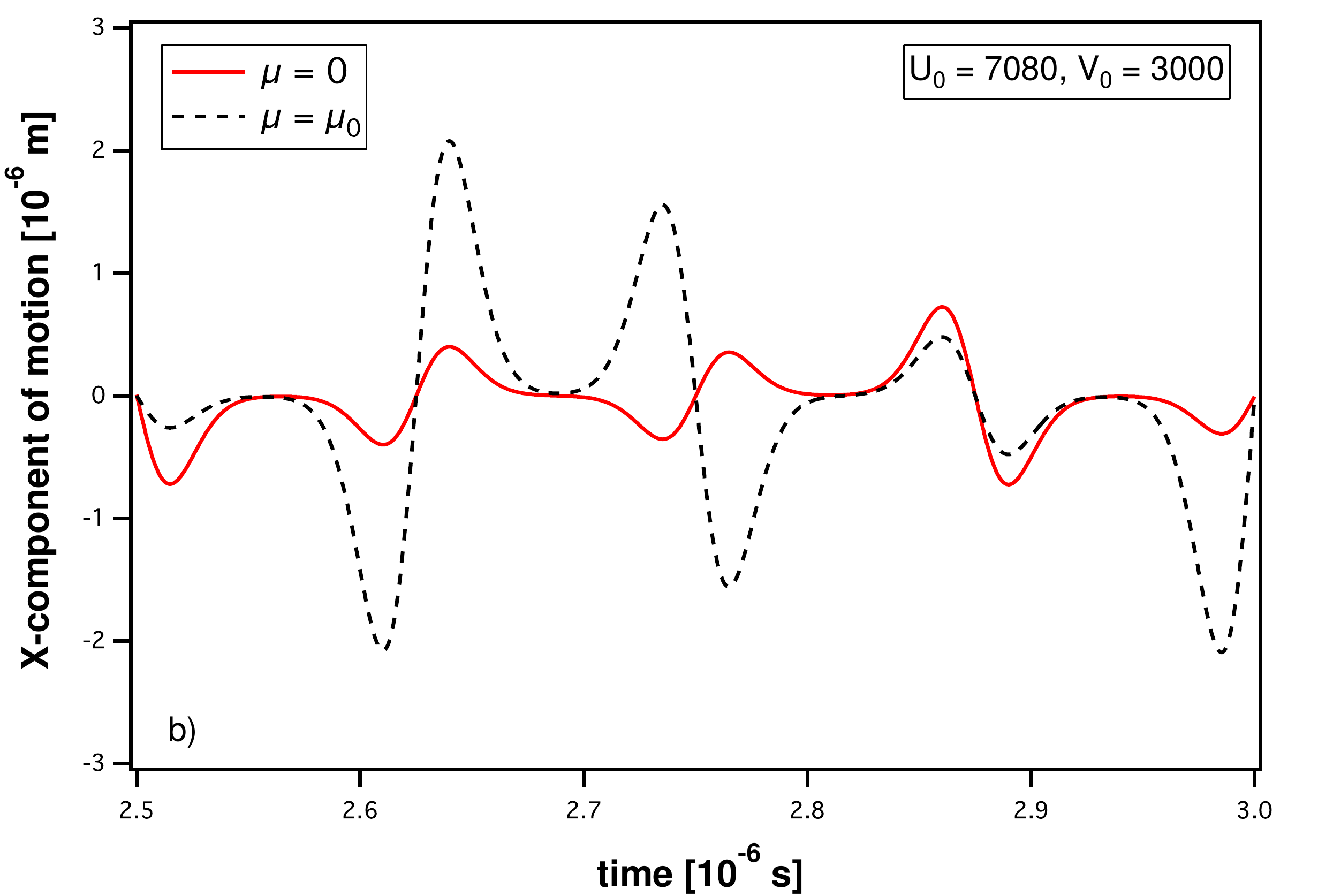}
\caption{(Color online) Time evolution of the $X$ component of the
  center-of-mass motion, with classical rotation, for different trap
  potentials: (a) $U_{0} = \SI{2175}{\volt}$, $V_{0} =
  \SI{2000}{\volt}$; (b) $U_{0} = \SI{7080}{\volt}$, $V_{0} =
  \SI{3000}{\volt}$.}
\label{fig:class_mu_vs_Z}
\end{figure}

Increasing the magnitude of the fields $(U_{0}, V_{0})$ increases the
magnitude of the interaction energy between the dipole moment and the
electric field.  Results for the previous field $(U_{0} =
\SI{680}{\volt}, V_{0} = \SI{1650}{\volt})$ are compared to stronger
fields in Fig.~\ref{fig:class_second}, where we plot the difference in
trajectories with and without a dipole moment, only for the $X$
component of motion.  To increase visibility, we have only plotted a
part of the entire trajectory, in the time interval
$[\SI{2.5}{\micro\second}, \SI{3.0}{\micro\second}]$.
We see in Fig.~\ref{fig:class_second} that, as expected, the deviation
of the trajectories increases with the magnitude of the trapping
fields.  In the absence of a dipole moment, the trajectories are only
moderately affected by the change in the trapping field, while the
increased coupling with the dipole moment in the case of the stronger
field results in a significant change of trajectory, as evidenced in
Fig.~\ref{fig:class_mu_vs_Z}.

\begin{figure}
\centering
\includegraphics[scale=0.50]{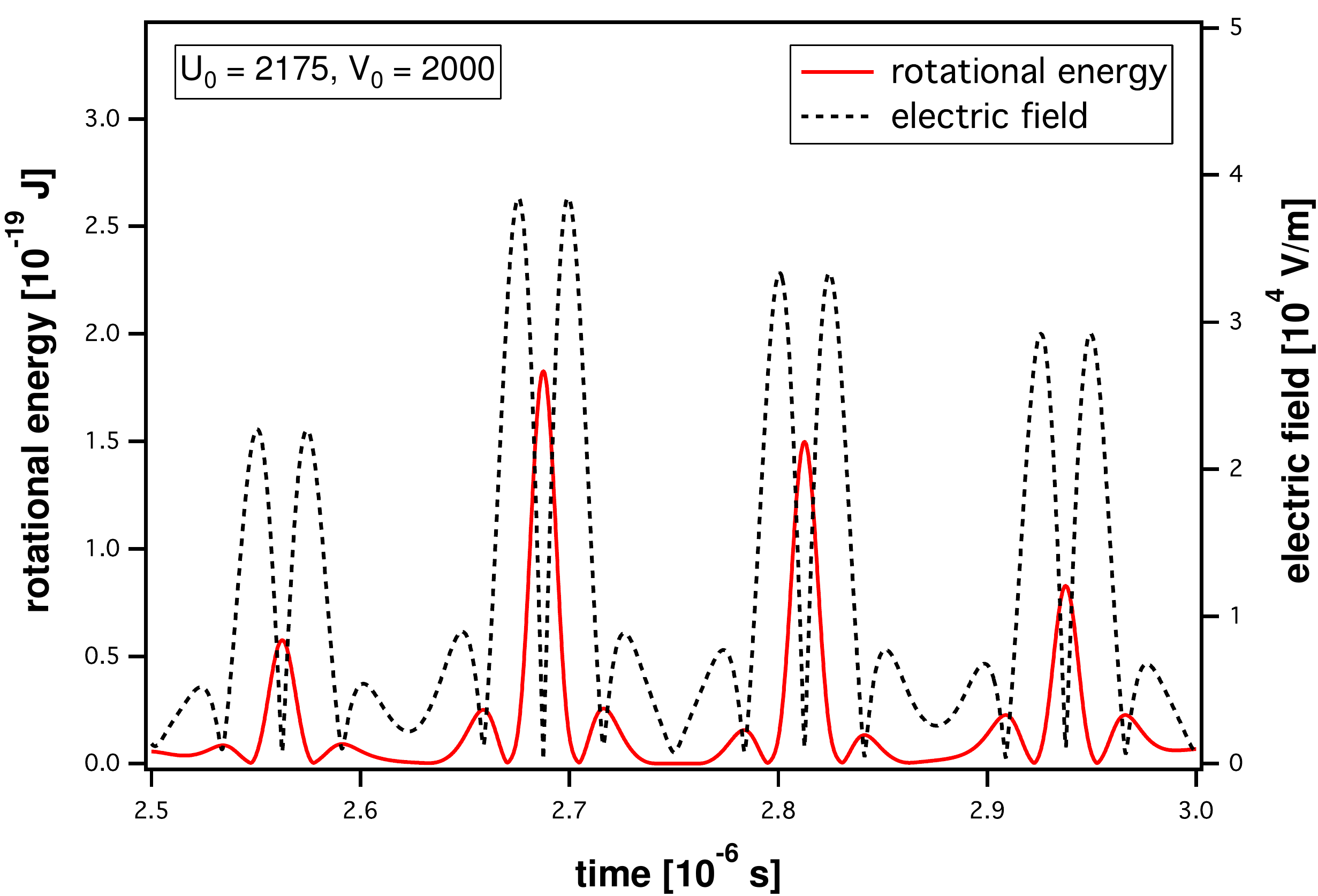}
\caption{(Color online) Time evolution of the rotational energy of a
  classical $\mathrm{MgH}^{+}$ molecular ion in a linear Paul
  trap. The time evolution of the trapping electric field, at the
  position $\mathbf{R}$ of the ion, is also shown as a dotted (black)
  curve. The magnitudes of the trapping potentials are $U_{0} =
  \SI{2175}{\volt}, V_{0} = \SI{2000}{\volt}$.}
\label{fig:class_rotE}
\end{figure}

We have also investigated the role of the initial orientation of the
ion, by changing the initial values of $\theta$ and $\varphi$, as
mentioned above.  The result is also shown in
Fig.~\ref{fig:class_second}, where we see a strong dependence of the
trajectory on this initial orientation.  It is interesting to note
that, nevertheless, the deviations in the trajectory show the same
periodicity and have the same zero crossings. It appears that the
orientation of the dipole significantly changes the trajectory only at
the turning points of the latter.  As expected, these turning points
are also where the rotations are the most hindered by the
dipole-field coupling, as evidenced by
the rotational energy
\begin{equation}
E_{\mathrm{rot}} = \frac{1}{2}I(\omega_{P{x}'}^{2} + \omega_{P{y}'}^{2})\, ,
\end{equation}
see Fig.~\ref{fig:class_rotE}.

\subsection{Quantum rotation}
\label{subsec:res_semi_class}

In the semi-classical simulations, the initial state of the ion is
chosen as for the classical simulations, except for rotations, where
we take the initial quantum rotational state to correspond to a
specific spherical harmonic $\ket{J,M}$.  In other words, all
coefficients $D_{J,M}$ in Eq.~(\ref{eq:rot_function}) are equal to
zero at $t=0$, except for one which is set to $1$.  All the results
presented here for the semi-classical model are for the highest
magnitude of the trapping field used in the classical simulations,
namely $U_{0} = \SI{7080}{\volt}$, $V_{0} = \SI{3000}{\volt}$.

\subsubsection{Effect of the dipole moment on the trajectory}

\begin{figure}
\begin{minipage}[c]{0.45\linewidth}
\includegraphics[scale=0.33]{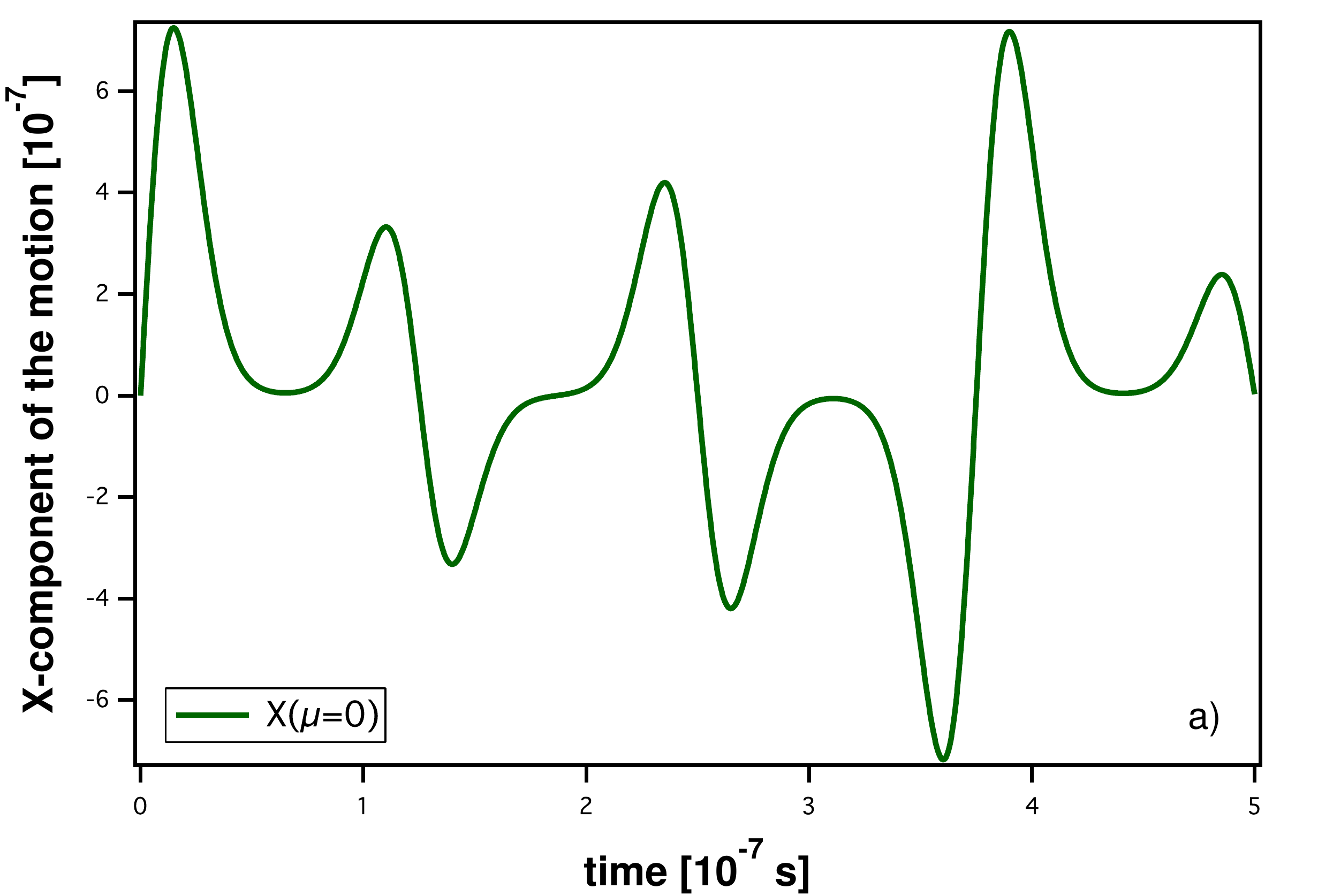}
\end{minipage}
\hspace{0.5cm}
\begin{minipage}[c]{0.50\linewidth}
\includegraphics[scale=0.33]{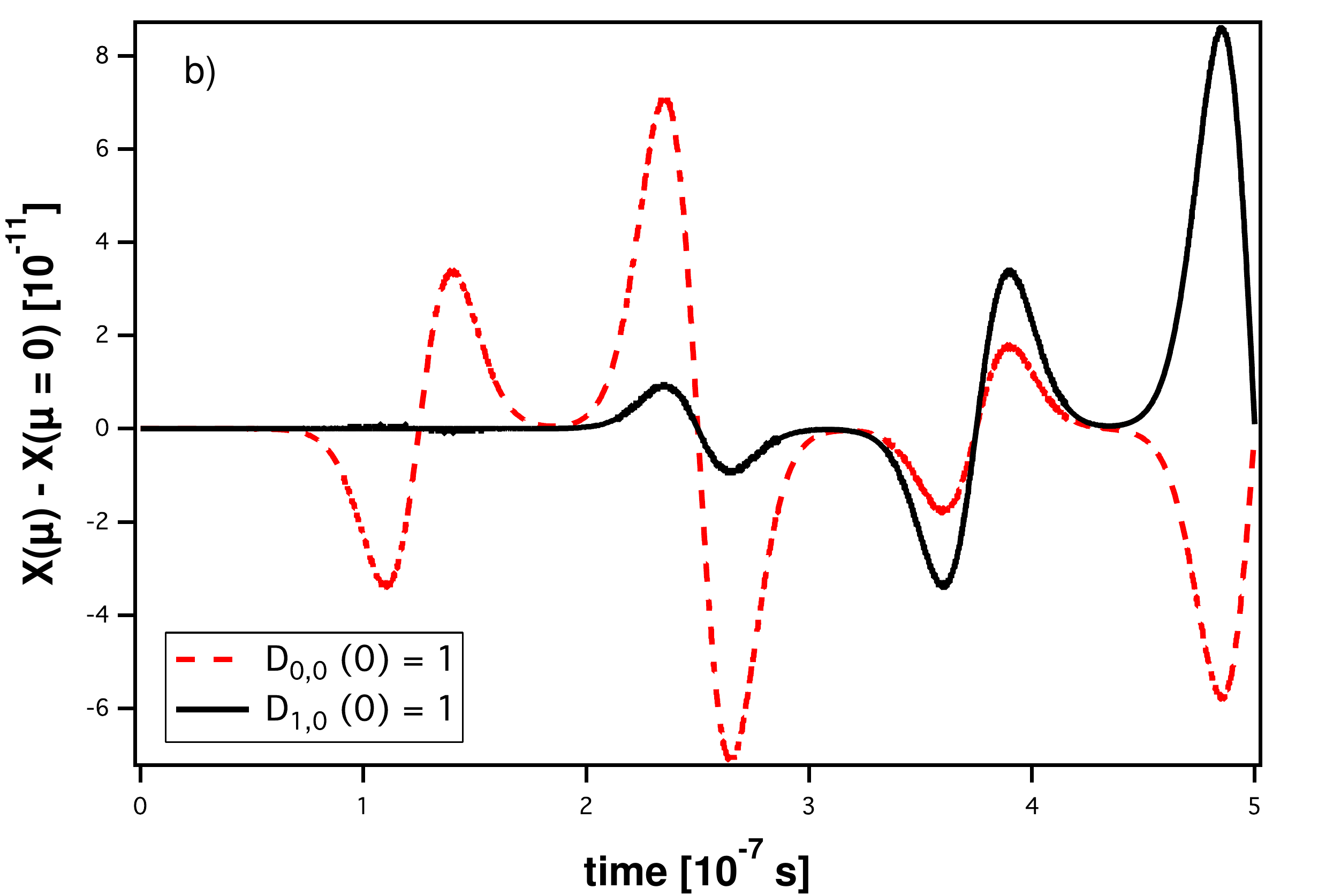}
\end{minipage}
\begin{minipage}[c]{0.45\linewidth}
\includegraphics[scale=0.33]{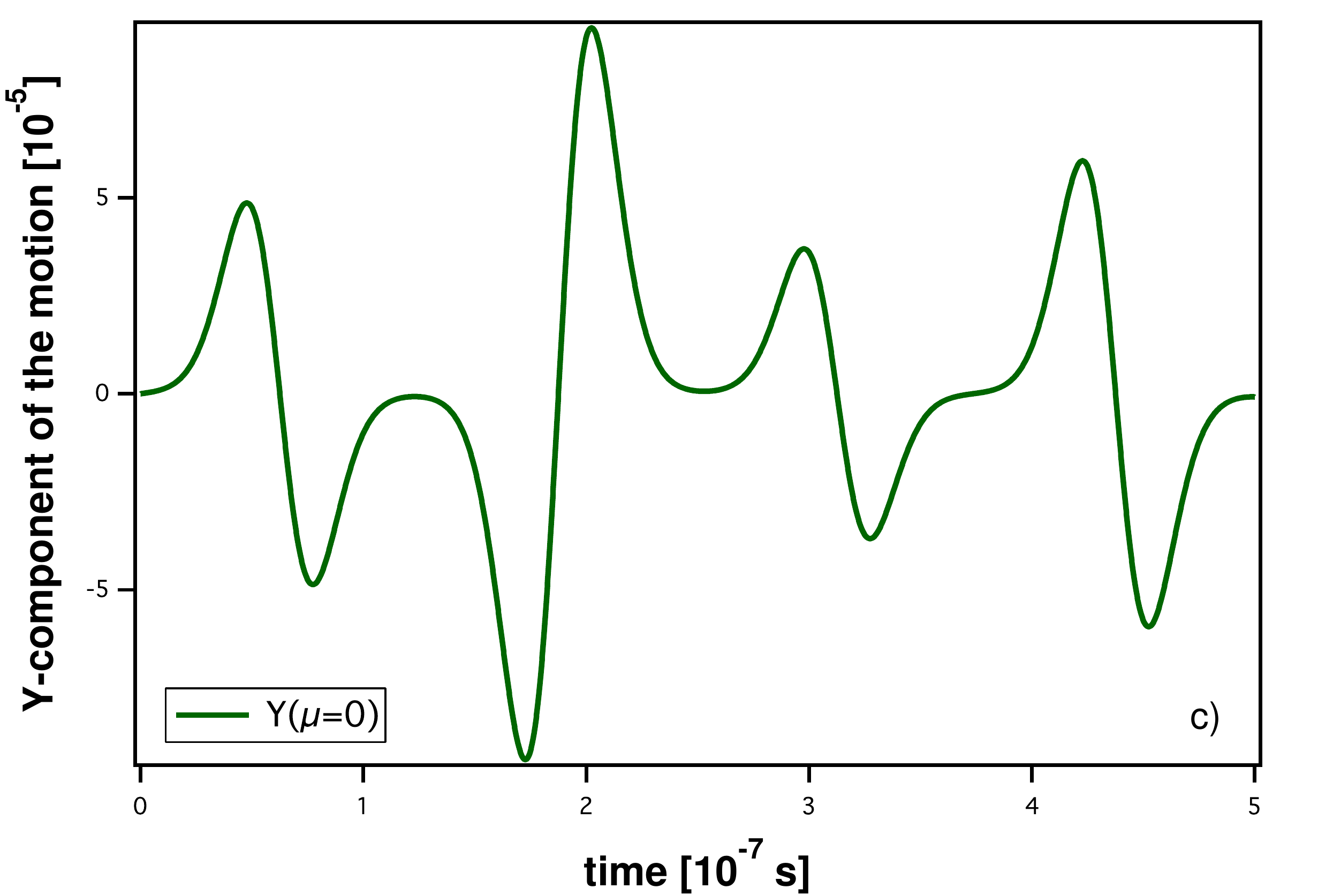}
\end{minipage}
\hspace{0.5cm}
\begin{minipage}[c]{0.5\linewidth}
\includegraphics[scale=0.33]{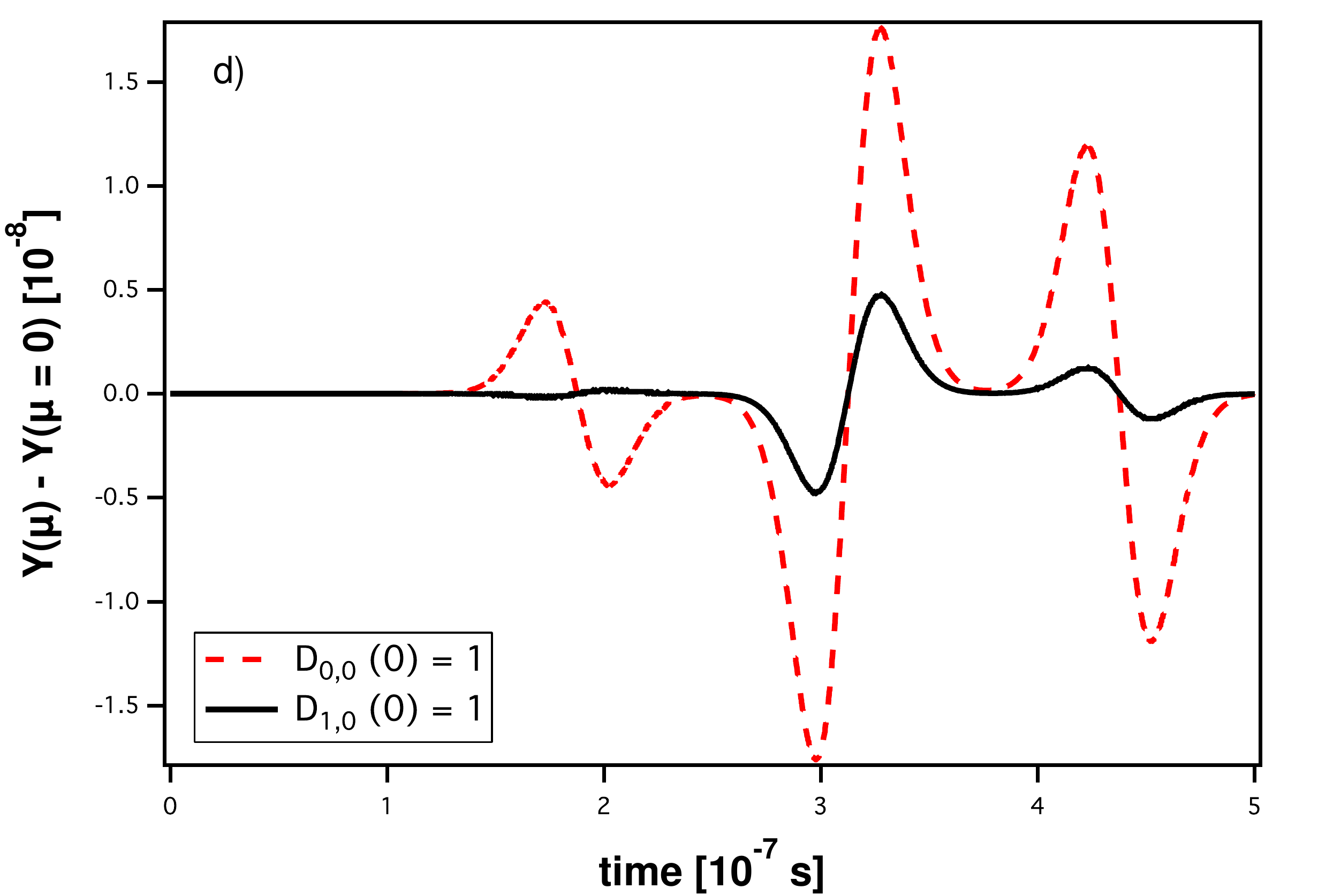}
\end{minipage}
\caption{(Color online) Time evolution of the center-of-mass motion of
  a $\mathrm{MgH^{+}}$ ion inside a linear Paul trap, with quantum
  rotation, for: (a)--(b) $X$ component of motion, (c)--(d)
  $Y$ component of motion. Panels (a) and (c) show the trajectories of
  the motion when $\mu = 0$. Panels (b) and (d) present the difference
  between the trajectories of motion when $\mu = 0$ and when $\mu \neq
  0$. In the case when $\mu \neq 0$, the ion is initially in a
  specific rotational state.}
\label{fig:semi_class_traj}
\end{figure}

The results for the center-of-mass motion of the ion in the
semi-classical model, starting from the initial rotational states
$\ket{0,0}$ and $\ket{1,0}$, are shown in
Fig.~\ref{fig:semi_class_traj}.  As previously, we also present the
difference between the trajectories obtained with and without a dipole
moment.  Only results along the $X$ and $Y$ axes are presented, as
there is no effect on the $Z$ component of the trajectory.
Comparing to the result for the classical model,
Fig.~\ref{fig:class_second}, we see that the effect of the quantum
rotation on the center-of-mass motion is smaller.  
%
%
In addition, the
deviations in the trajectories are smaller when the ion is
initially in an excited state ($J = 1$) in comparison to the
rotational ground state ($J = 0$).

\subsubsection{Effect of the trapping field on the rotation of the ion}

\begin{figure}
\centering
\includegraphics[scale=0.50]{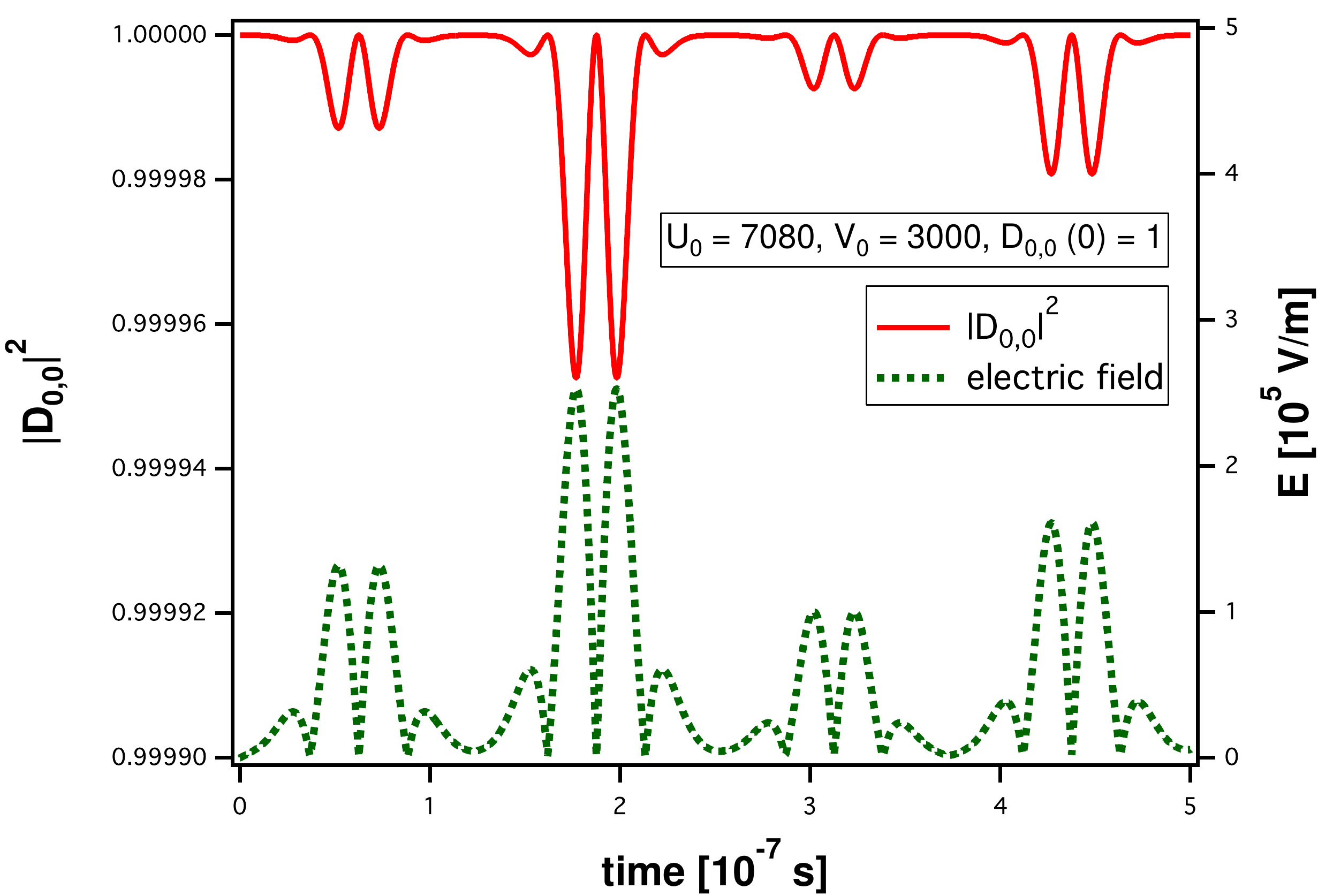}
\caption{(Color online) Time evolution of the quantum rotational
  ground state of a $\mathrm{MgH}^{+}$ molecular ion with permanent
  dipole moment $\mu_{0}$, in a linear Paul trap. The ion is initially
  in its rotational ground state. The time evolution of the trapping
  electric field, at the position $\mathbf{R}$ of the ion, is also
  shown with a dotted (green) curve. The magnitudes of the trapping
  potentials are $U_{0} = \SI{7080}{\volt}, V_{0} =
  \SI{3000}{\volt}$.}
\label{fig:semi_class_first}
\end{figure}

We now shift our attention from the center-of-mass motion to the
effect of the external trapping electric field on the rotational
dynamics of the molecular ion.  In Fig.~\ref{fig:semi_class_first}, we
plot the time evolution of $\left | D_{0,0}(t) \right |^{2}$, the
probability of finding the ion in the ground rotational state
$\ket{0,0}$.  
We see that the trapping field causes a small perturbation in the
quantum rotation of the diatomic ion, as is clearly evidenced by the
local value of the field felt by the ion, which is also plotted in
Fig.~\ref{fig:semi_class_first}.  The trapping field is thus coupling
the ground rotational state to excited states, but the ion appears to
return adiabatically to the ground state as it revisits a low-field
region.

\begin{figure}
\centering
\includegraphics[scale=0.33]{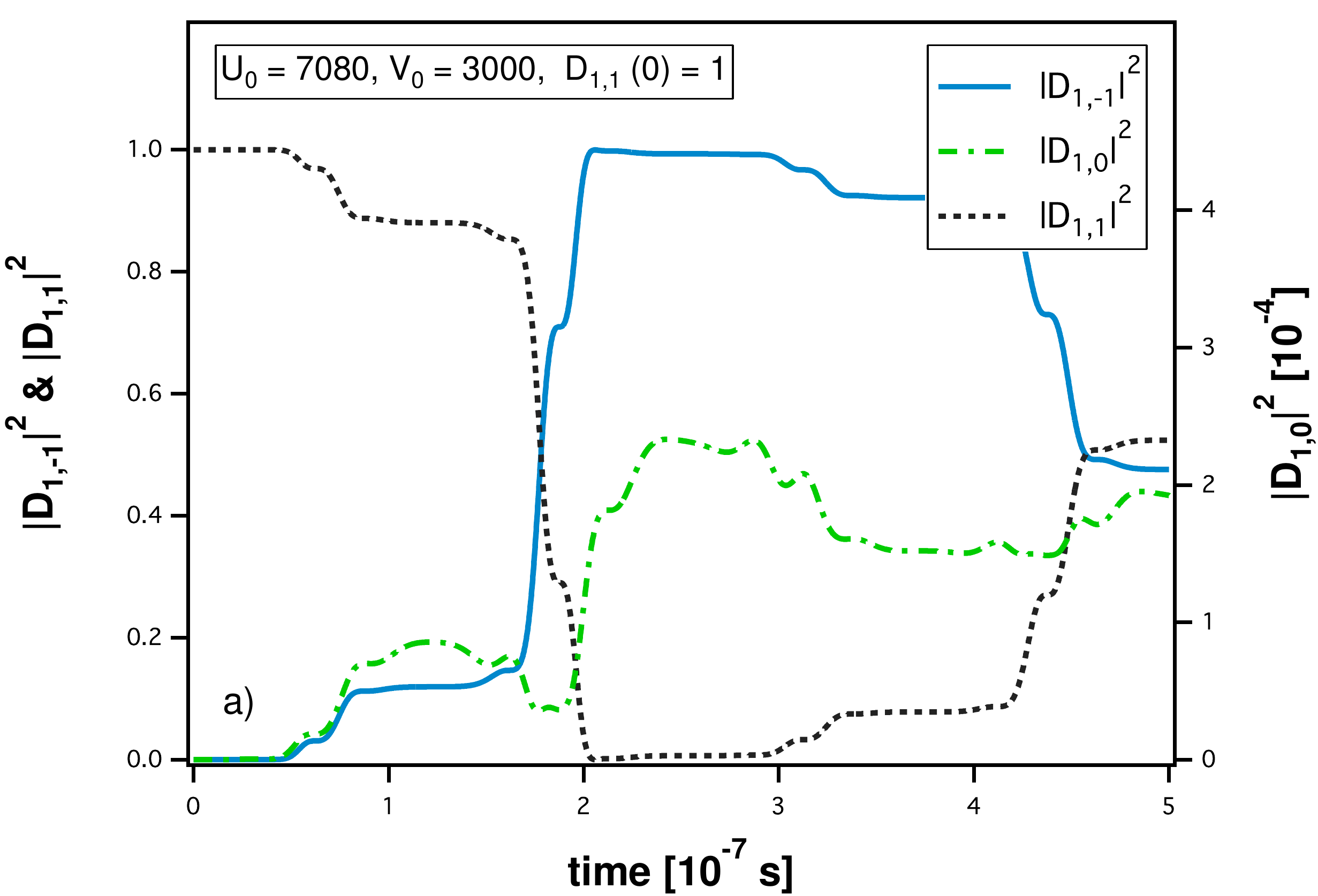}
\centering
\includegraphics[scale=0.33]{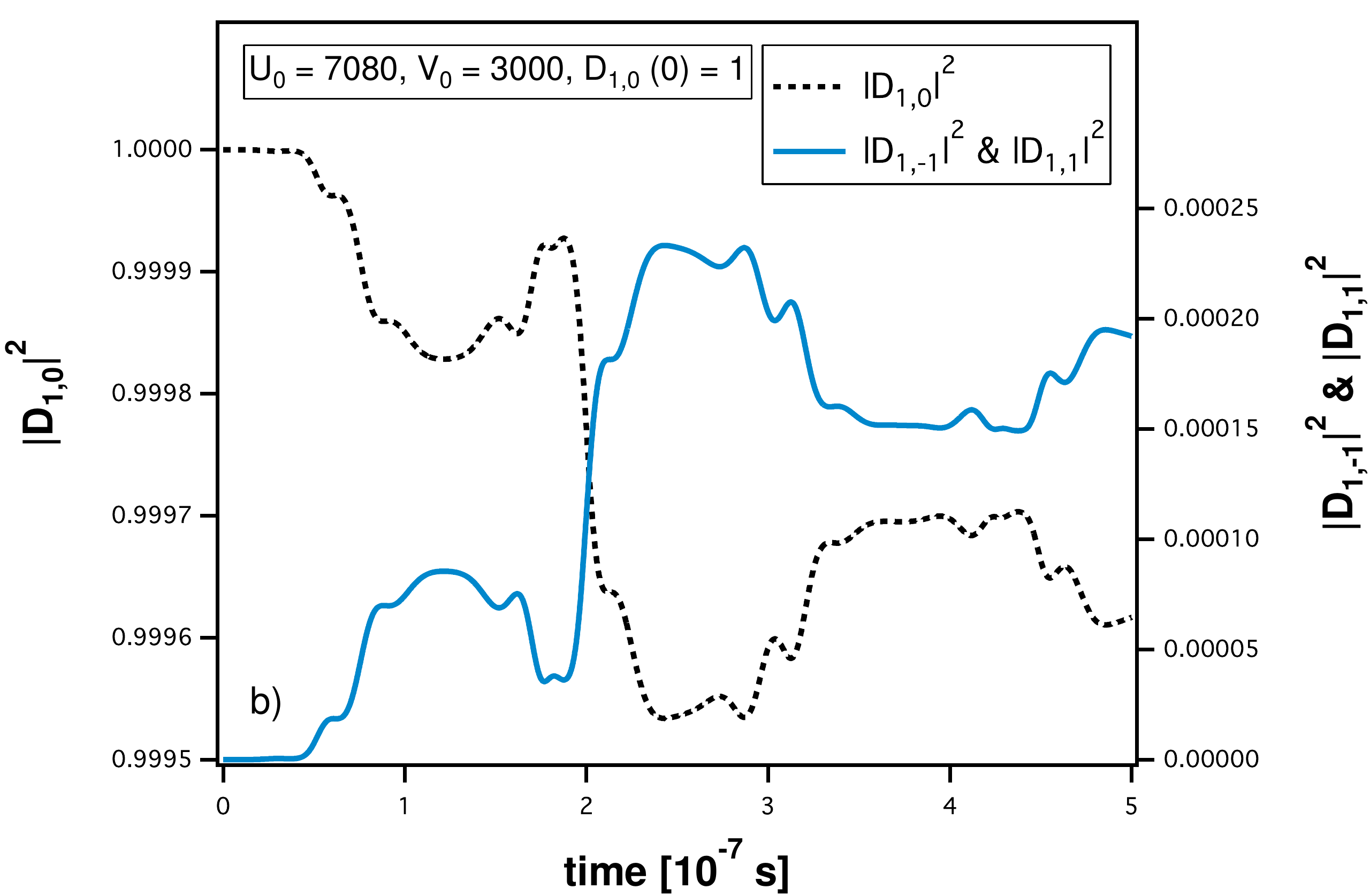}
\centering
\includegraphics[scale=0.33]{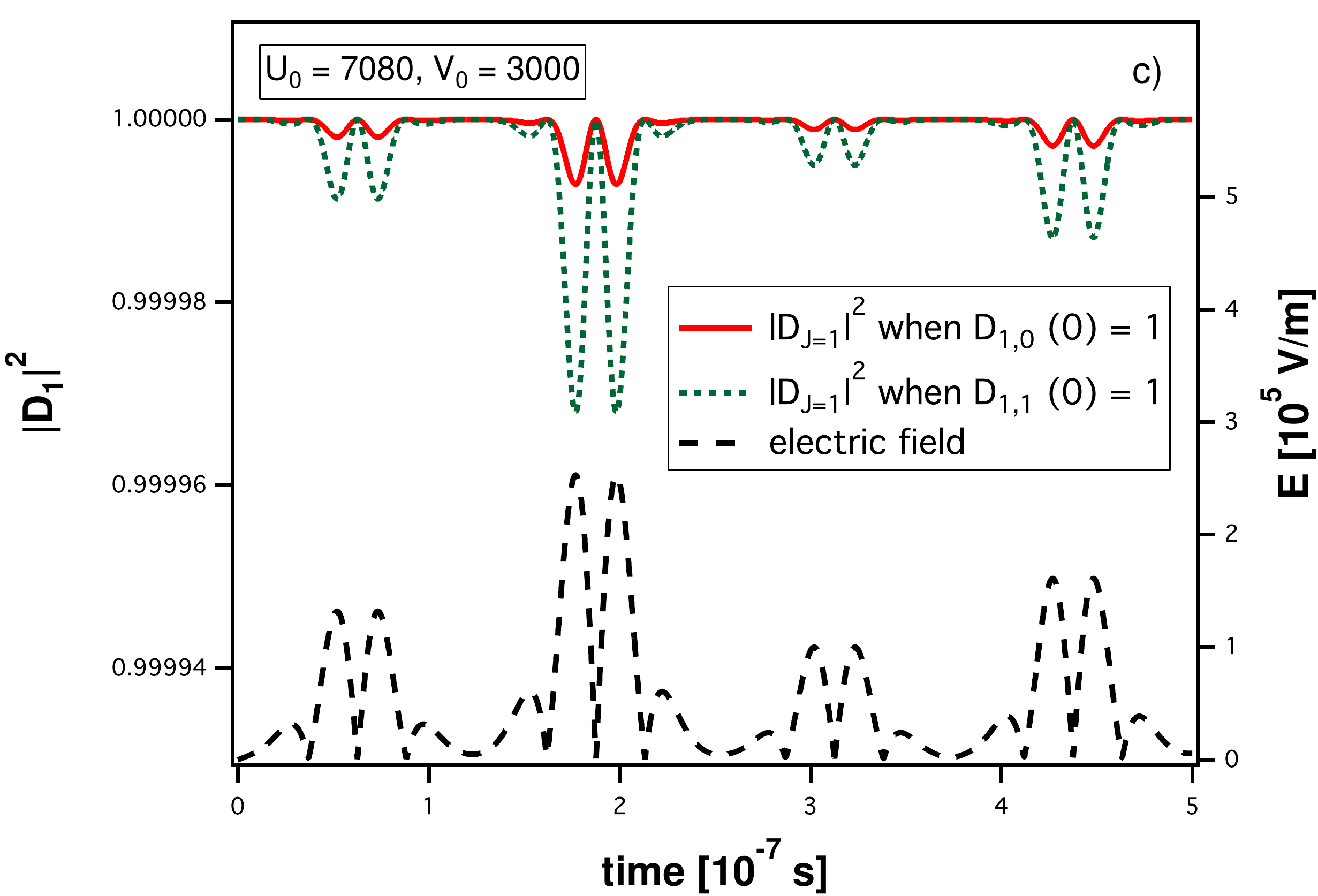}
\caption{(Color online) Time evolution of the first excited quantum
  rotational state of a $\mathrm{MgH}^{+}$ molecular ion with
  permanent dipole moment $\mu_{0}$, in a linear Paul trap. Panels
  (a)--(b) show the time evolution of sub-levels $M = 0, \pm 1$ when
  the ion is initially in either rotational states corresponding to $J
  = 1$, $M = 1$ [panel (a)] and $J = 1$, $M = 0$ [panel(b)]. Panel (c)
  compares the time evolution of state with $J = 1$, for the two initial
  conditions mentioned, along with the time evolution of the trapping
  electric field at the position $\mathbf{R}$ of the ion.  The
  magnitudes of the trapping potentials are $U_{0} = \SI{7080}{\volt},
  V_{0} = \SI{3000}{\volt}$.}
\label{fig:semi_class_second}
\end{figure}

Comparable results are obtained when starting from the excited states
$\ket{1,0}$ or $\ket{1,1}$, as shown in
Fig.~\ref{fig:semi_class_second}.  
Starting from the state $\ket{1,1}$, one sees,
Fig.~\ref{fig:semi_class_second}(a), a strong exchange with the
corresponding sub-level with opposite $M=-1$, $\ket{1,-1}$, with
limited transfer to the state $\ket{1,0}$ [note the difference in
scale in Fig.~\ref{fig:semi_class_second}(a) for the latter state].
Similarly, starting from $\ket{1,0}$, the ion basically stays in that
state, as shown in Fig.~\ref{fig:semi_class_second}(b).

By summing over the probabilities of finding the ion to be in one of
the sub-levels above, \ie,
\begin{equation}
  \left| D_{J}(t) \right |^{2} \equiv \sum_{M = - J}^{J} \left | D_{J,M}(t)
  \right |^{2}
  \label{eq:sublevels}
\end{equation}
for $J=1$, we see in Fig.~\ref{fig:semi_class_second}(c) that the
excited states are less perturbed by the external field than the ground
state, especially when starting from $\ket{1,0}$.  Nevertheless, the
states corresponding to $J = 1$ follow the same temporal changes as
the local electric field, as was the case for the ground state.

\begin{figure}
\centering
\includegraphics[scale=0.50]{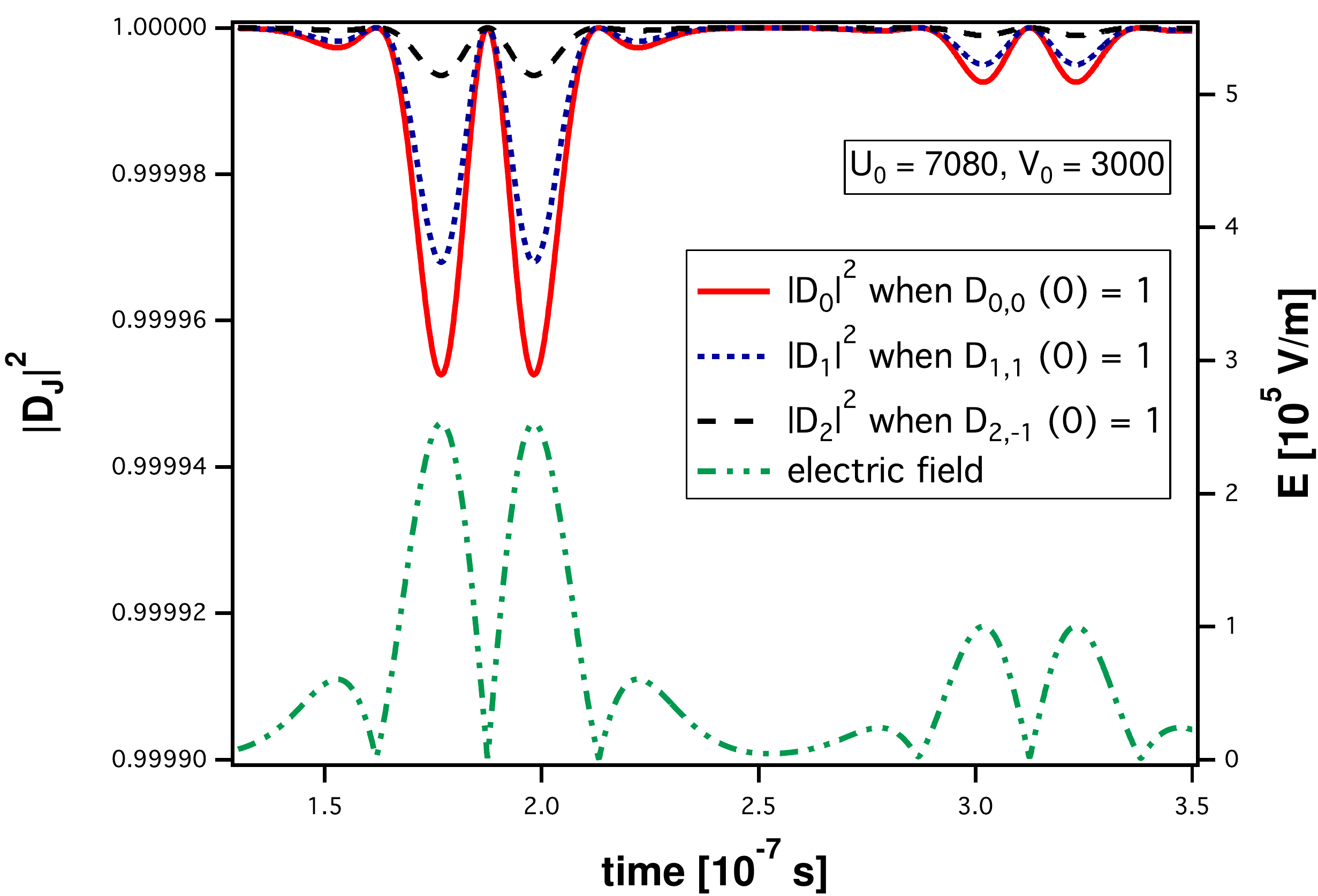}
\caption{(Color online) Comparison of the time evolution of the
  quantum rotational ground state $J = 0$ (when the ion is initially
  in this state), first rotational excited state $J = 1$ (when the ion
  is initially in this state), and second rotational excited state $J
  = 2$ (when the ion is initially in this state) with each other and
  with the time evolution of the trapping electric field, at the
  position $\mathbf{R}$ of the ion. The magnitudes of the trapping
  potentials are $U_{0} = \SI{7080}{\volt}, V_{0} =
  \SI{3000}{\volt}$.}
\label{fig:semi_class_forth}
\end{figure}

Finally, we have also compared the time evolution of the first three
rotational states of the ion corresponding to quantum numbers $J = 0$,
$J = 1$ and $J = 2$, by studying their populations $\left
  | D_{J}(t) \right |^{2}$, Eq.~(\ref{eq:sublevels}), in
Fig.~\ref{fig:semi_class_forth}.
As we see, the rotational ground state is the most affected state by
the interaction of the electric field with the dipole, while on the
other hand, the states corresponding to $J = 2$ are hardly affected by
the trapping field. This confirms the tendency that was seen in
Figs.~\ref{fig:semi_class_first} and \ref{fig:semi_class_second}, that
the higher the energy of the rotational state, the less it is affected
by the interaction between the trapping electric field and the dipole
moment.  It can be understood by considering that the lower the value
of $J$, the more the presence of the dipole-field interaction
$\bm{\mu}\cdot\mathbf{E}$ affects the rotational
eigenstates, see Refs.~\onlinecite{vonMeyenn_ZP_1970,Friedrich_ZPD_1991}.

\section{Conclusion}
\label{sec:con}

We have studied the dynamics of a rigid heteronuclear diatomic
molecular ion trapped in a linear Paul trap by carrying out both
classical and semi-classical numerical simulations. We have
investigated the effect of the interaction between the permanent
dipole moment of the ion and the time-varying trapping electric field
on the motion of the center of mass of the ion, along with the effect
of the field on the rotation of the ion.

Classically, we showed that the trajectories of the center of mass
differ from those of an equivalent ion without a permanent dipole
moment.  The trajectories obtained therefore do not follow the Mathieu
equation, as atomic ions do, but present nevertheless the same
periodic oscillations.  We observed that the deviation of the
trajectories of the motion of the molecular ion is strongly dependent
on its initial orientation (orientation of the dipole) inside the
trap, for the same magnitude of the electric field.  Similarly, a
stronger trapping field lead to a greater divergence between the
trajectories with and without a coupling between the field and the
dipole moment of the ion.

Considering quantum rotation in a semi-classical model, we found that
the time evolution of the rotational states of the ion follows closely
the time evolution of the trapping electric field. Therefore, when the
electric field is zero, the initial rotational state of the ion is
fully populated and as soon as the electric field increases from zero,
the probability of finding the ion in other rotational states becomes
non-zero.  We showed that higher rotational states of the ion are less
affected by the interaction of the dipole moment with the trapping
electric field. We also found that the trajectory of the ion is less
affected by the coupling between the trapping field and the dipole
moment when quantum rotation is considered, compared to classical rotation.

In all cases, we found that the deviations of the trajectories from
the Mathieu equation did not lead to unbounded trajectories.  In other
words, the stability of the trap was not affected by the presence of
the coupling between the dipole moment and the trapping field for the
potential values $U_0$ and $V_0$ used in this study.  Future work should
explore whether there exists instances where this is not the case,
maybe very close to the border between stable and unstable trajectories.

\begin{acknowledgments}
  This research was conducted using the resources of the High
  Performance Computing Center North (HPC2N).  Financial support from
  Ume{\aa} University is gratefully acknowledged.
\end{acknowledgments}

\appendix
\section{Classical model for a dipolar rigid rotor in an external
  electric field}
\label{app:model}

\subsection{Definition of the model}

\begin{figure}
\centering
\includegraphics[scale=0.5]{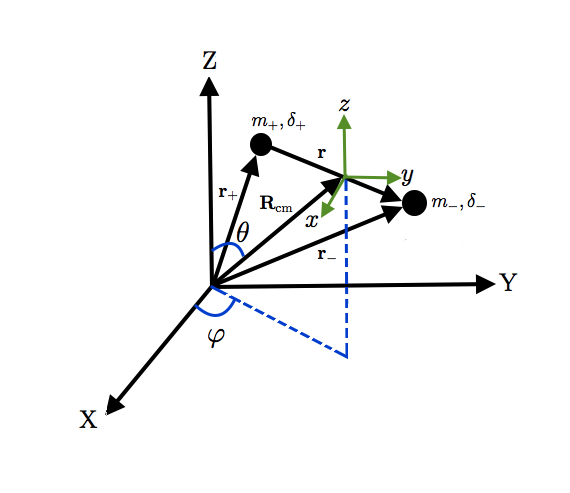}
\caption{(Color online) Representation of a dipolar, diatomic
  molecular ion as a system of two masses $m_{+}$ and $m_{-}$
  possessing charges $\delta_{+}$ and $\delta_{-}$, respectively.}
\label{fig:model}
\end{figure}

We represent a dipolar diatomic molecular ion as a classical rigid
rotor, with two partial charges $\delta_+$ and $\delta_-$ located at
$\mathbf{r}_+$ and $\mathbf{r}_-$, see Fig.~\ref{fig:model}.
In order to reproduce the characteristics of the molecular ion, $r
\equiv \left| \mathbf{r} \right|$ is fixed to the internuclear
distance, with $\mathbf{r} \equiv \mathbf{r}_+ - \mathbf{r}_-$, and
the two charges are assigned the masses $m_+$ and $m_-$ of the
corresponding atoms.  The position of the center of mass is given by
\begin{equation}
\mathbf{R} = \frac{m_+ \mathbf{r}_+ + m_- \mathbf{r}_-}{m_+ + m_-} = \frac{m_+
  \mathbf{r}_+ + m_- \mathbf{r}_-}{M},
\end{equation}
where $M \equiv m_+ + m_-$ is the total mass of the molecular ion.
The position of the partial charges relative to the center of mass,
$\mathbf{r}'_\pm$, are obtained from
\begin{equation}
\label{eq:re_pos}
  \mathbf{r}_\pm = \mathbf{R} + \mathbf{r}'_\pm,
\end{equation}
from which we get 
\begin{equation}
  m_+ \mathbf{r}_+ + m_- \mathbf{r}_- = M \mathbf{R} + m_+ \mathbf{r}_+' + m_- \mathbf{r}_-',
\end{equation}
so by definition of the center of mass, we have that
\begin{equation}
  m_+ \mathbf{r}_+' + m_- \mathbf{r}_-' = 0.
\end{equation}
Similarly, starting from the momentum
\begin{equation}
  m_+ \dot{\mathbf{r}}_+ + m_- \dot{\mathbf{r}}_- = M \dot{\mathbf{R}},
\end{equation}
we deduce that
\begin{equation}
  m_+ \dot{\mathbf{r}}_+' + m_- \dot{\mathbf{r}}_-' = 0
  \label{eq:relative_momentum}
\end{equation}

Using the above definitions, we can find the partial charges by
imposing that
\begin{equation}
  \delta_+ + \delta_- = Ze,
\label{eq:total_charge}
\end{equation}
with $Ze$ the total charge of the ion and $e$ the elementary charge,
along with
\begin{equation}
  \delta_+ \mathbf{r}_+' + \delta_- \mathbf{r}_-' = \bm{\mu},
\label{eq:dipole}
\end{equation}
with $\bm{\mu}$ the dipole moment of the molecular ion.

\subsubsection{Center-of-mass frame}

We define an additional reference frame $(x, y, z)$ centered on the
center of mass of the molecular ion, that remains aligned with the lab
reference frame $(X, Y, Z)$, \ie, $\mathbf{\hat{x}}$ is always parallel to
$\mathbf{\hat{X}}$, etc., where the $\mathbf{\hat{\mbox{ }}}$ denotes unit vectors.  In
the center-of-mass frame, the vector $\mathbf{r}_+$ can be expressed
in spherical coordinates as
\begin{equation}
  \mathbf{r}_+ = \mathbf{\hat{x}} r_+' \sin \theta \cos \varphi + \mathbf{\hat{y}} r_+' \sin
  \theta \sin \varphi + \mathbf{\hat{z}} r_+' \cos \theta,
\end{equation}
with its time derivative 
\begin{align}
  \dot{\mathbf{r}}_+ &= \mathbf{\hat{x}} r_+' \left( \dot{\theta} \cos \theta
    \cos \varphi - \dot{\varphi} \sin \theta \sin \varphi \right) \nonumber \\
  & \quad +  \mathbf{\hat{y}} r_+' \left( \dot{\theta} \cos \theta
    \sin \varphi + \dot{\varphi} \sin \theta \cos \varphi \right) \nonumber \\
  & \quad +  \mathbf{\hat{z}} r_+' \left(- \dot{\theta} \sin \theta \right),
\end{align}
where we have made use of the rigid rotor approximation, $\dot{r}_+' =
0$.  From this, we get
\begin{align}
  \dot{\mathbf{r}}_+ \cdot \dot{\mathbf{r}}_+ &= \left(r_+'\right)^2
  \left[ \left( \dot{\theta} \cos \theta \cos \varphi - \dot{\varphi}
      \sin \theta \sin \varphi \right)^2
  \right. \nonumber \\
  & \quad \quad \quad + \left( \dot{\theta} \cos \theta \sin \varphi +
    \dot{\varphi} \sin \theta \cos \varphi \right)^2 \nonumber
  \\
  & \quad \quad \quad \left. + \left(- \dot{\theta} \sin \theta
    \right)^2
  \right], \nonumber \\
  \intertext{which, after some simple algebra, results in} &=
  \left(r_+'\right)^2 \left[ \dot{\theta}^2 + \dot{\varphi}^2
    \sin^2\theta \right].
  \label{eq:vp2}
\end{align}
For $\mathbf{r}_-$, we have to make the substitutions $\theta
\rightarrow \pi - \theta$ and $\varphi \rightarrow \varphi + \pi$,
hence
\begin{equation}
  \mathbf{r}_- = - \mathbf{\hat{x}} r_-' \sin \theta \cos \varphi - \mathbf{\hat{y}} r_-' \sin
  \theta \sin \varphi - \mathbf{\hat{z}} r_-' \cos \theta,
\end{equation}
from which we recover
\begin{equation}
  \dot{\mathbf{r}}_- \cdot \dot{\mathbf{r}}_- = \left(r_-'\right)^2
  \left[  \dot{\theta}^2  + \dot{\varphi}^2 \sin^2\theta \right]. 
  \label{eq:vm2}
\end{equation}

\subsection{Kinetic energy}

Using the notation $\mathbf{v} \equiv \dot{\mathbf{r}}$, the kinetic energy
is given by
\begin{align}
  K &= \frac{1}{2} \left( m_+ v_+^2 + m_- v_-^2 \right) \nonumber \\
  &= \frac{1}{2} \left( m_+ \mathbf{v}_+ \cdot \mathbf{v}_+ + m_- \mathbf{v}_-
    \cdot \mathbf{v}_- \right) \nonumber \\
  &= \frac{1}{2} \left[ m_+ \left( \dot{\mathbf{R}} \cdot \dot{\mathbf{R}} +
      2 \dot{\mathbf{R}} \cdot \dot{\mathbf{r}}_+' + \dot{\mathbf{r}}_+' \cdot
      \dot{\mathbf{r}}_+' \right) + m_- \left( \dot{\mathbf{R}} \cdot
      \dot{\mathbf{R}} + 2 \dot{\mathbf{R}} \cdot \dot{\mathbf{r}}_-' +
      \dot{\mathbf{r}}_-' \cdot
      \dot{\mathbf{r}}_-' \right) \right] \nonumber \\
  &= \frac{1}{2} \left[ M V^2 + 2 \dot{\mathbf{R}} \cdot \left( m_+
      \dot{\mathbf{r}}_+' + m_- \dot{\mathbf{r}}_-' \right) + m_+ v_+'^2 +
    m_- v_-'^2 \right] \nonumber \\
  &= \frac{1}{2} \left( M V^2 + m_+ v_+'^2 + m_- v_-'^2 \right),
\end{align}
where we have used Eq.~(\ref{eq:relative_momentum}).  Using
Eqs.~(\ref{eq:vp2}) and (\ref{eq:vm2}), we can express the kinetic
energy in spherical coordinates as
\begin{align}
  K &= \frac{1}{2} \left( M \dot{\mathbf{R}}^2 + m_+ \dot{\mathbf{r}}_+'^2 +
    m_- \dot{\mathbf{r}}_-'^2 \right) \nonumber \\
  &= \frac{1}{2} M \dot{\mathbf{R}}^2 + \frac{1}{2} \left( m_+ r_+'^2 +
    m_- r_-'^2 \right) \left( \dot{\theta}^2
    + \dot{\varphi}^2 \sin^2\theta \right).
\end{align}
From the definition of the center of mass, we have
\begin{align}
  m_+ r_+'^2 + m_- r_-'^2 &= m_+ \left( \frac{m_-}{M} r \right)^2 + m_-
  \left( \frac{m_+}{M} r \right)^2 \nonumber \\
  &= \frac{m_+ m_-^2}{M^2} r^2 + \frac{m_- m_+^2}{M^2} r^2 \nonumber
  \\
  &= \frac{m_+ m_-}{M} \left( \frac{m_+ + m_-}{M} \right) r^2
  \nonumber \\
  &= m_\mathrm{red} r^2
\end{align}
where the reduced mass $m_\mathrm{red} \equiv m_+ m_- / M$.  We thus get
\begin{align}
  K &= \frac{1}{2} M \dot{\mathbf{R}}^2 + \frac{1}{2} m_\mathrm{red} r^2
  \left( \dot{\theta}^2 + \dot{\varphi}^2 \sin^2\theta \right) \nonumber
  \\
  &= \frac{1}{2} M \dot{\mathbf{R}}^2 + \frac{1}{2} I
  \dot{\bm{\omega}}^2,
\end{align}
where $I \equiv  m_\mathrm{red} r^2$ is the moment of inertia and
$\dot{\bm{\omega}}$ the angular velocity.

\subsection{Potential energy}
\label{subsec:pot_app}

We consider only the interaction of the partial charges $\delta_\pm$
with an external electric field $\Phi$,
\begin{equation}
V = \delta_+ \Phi(\mathbf{r}_+) + \delta_- \Phi(\mathbf{r}_-).
\end{equation}
Considering the form of the electric potential inside a Paul trap [see
Eqs.~(\ref{eq:tr_pot}) and (\ref{eq:tr_pots})], we can separate it
into Cartesian components
\begin{equation}
\Phi(X,Y,Z) = \Phi_X X^2 + \Phi_Y Y^2 + \Phi_Z Z^2 ,
\end{equation}
such that the Taylor expansion of $\Phi(\mathbf{r}_\pm) =
\Phi(\mathbf{R} + \mathbf{r}_\pm')$ around the center of mass is given
by
\begin{align}
  \Phi(\mathbf{R} + \mathbf{r}_\pm') &= \Phi(\mathbf{R}) + (\mathbf{r}_\pm')_{x}
  \left. \frac{\partial}{\partial x} \Phi \right|_{\mathbf{R}} + (\mathbf{r}_\pm')_{y}
  \left. \frac{\partial}{\partial y} \Phi \right|_{\mathbf{R}} + (\mathbf{r}_\pm')_{z}
  \left. \frac{\partial}{\partial z} \Phi \right|_{\mathbf{R}} \nonumber
  \\
  &\quad + (\mathbf{r}_\pm')_{x}^2
  \left. \frac{\partial^2}{\partial x^2} \Phi \right|_{\mathbf{R}} +
  (\mathbf{r}_\pm')_{y}^2 
  \left. \frac{\partial^2}{\partial y^2} \Phi \right|_{\mathbf{R}} +
  (\mathbf{r}_\pm')_{z}^2 
  \left. \frac{\partial^2}{\partial z^2} \Phi \right|_{\mathbf{R}}
\end{align}
with higher-order terms all zero. In deriving the previous equation,
we have made use of the fact that the coordinate systems $(X,Y,Z)$ and
$(x,y,z)$ are parallel, such that the operators $\partial / \partial
X$ and $\partial / \partial x$ are equivalent, etc.  We finally get
\begin{align}
\label{eq:app}
  V &= (\delta_+ + \delta_-) \Phi(\mathbf{R}) + \sum_{\xi = x,y,z} \left[
    \delta_+ (\mathbf{r}_+')_{\xi} + \delta_- (\mathbf{r}_-')_{\xi} \right]
    \left. \frac{\partial}{\partial \xi} \Phi \right|_{\mathbf{R}}
    \nonumber \\
    &\quad + \sum_{\xi = x,y,z} \left[
    \delta_+ (\mathbf{r}_+')_{\xi}^2 + \delta_- (\mathbf{r}_-')_{\xi}^2 \right]
    \left. \frac{\partial^2}{\partial \xi^2} \Phi \right|_{\mathbf{R}}
    \nonumber \\
    &= Z e \Phi(\mathbf{R}) + \bm{\mu} \cdot \left. \bm{\nabla}\Phi
    \right|_{\mathbf{R}} + \ldots \nonumber \\
    &\approx Z e \Phi(\mathbf{R}) - \bm{\mu} \cdot \mathbf{E}(\mathbf{R}),
\end{align}
where we have made use of Eqs.~(\ref{eq:total_charge}) and
(\ref{eq:dipole}), along with the relation for the electric field
$\mathbf{E} = - \nabla \Phi$.  Neglecting term including the second
derivative of the potential with respect to the coordinates, which is
the gradient of the field, corresponds to approximating the system as
a point dipole.

\bibliography{mol}

\begin{thebibliography}{46}%
\makeatletter
\providecommand \@ifxundefined [1]{%
 \@ifx{#1\undefined}
}%
\providecommand \@ifnum [1]{%
 \ifnum #1\expandafter \@firstoftwo
 \else \expandafter \@secondoftwo
 \fi
}%
\providecommand \@ifx [1]{%
 \ifx #1\expandafter \@firstoftwo
 \else \expandafter \@secondoftwo
 \fi
}%
\providecommand \natexlab [1]{#1}%
\providecommand \enquote  [1]{``#1''}%
\providecommand \bibnamefont  [1]{#1}%
\providecommand \bibfnamefont [1]{#1}%
\providecommand \citenamefont [1]{#1}%
\providecommand \href@noop [0]{\@secondoftwo}%
\providecommand \href [0]{\begingroup \@sanitize@url \@href}%
\providecommand \@href[1]{\@@startlink{#1}\@@href}%
\providecommand \@@href[1]{\endgroup#1\@@endlink}%
\providecommand \@sanitize@url [0]{\catcode `\\12\catcode `\$12\catcode
  `\&12\catcode `\#12\catcode `\^12\catcode `\_12\catcode `\%12\relax}%
\providecommand \@@startlink[1]{}%
\providecommand \@@endlink[0]{}%
\providecommand \url  [0]{\begingroup\@sanitize@url \@url }%
\providecommand \@url [1]{\endgroup\@href {#1}{\urlprefix }}%
\providecommand \urlprefix  [0]{URL }%
\providecommand \Eprint [0]{\href }%
\providecommand \doibase [0]{http://dx.doi.org/}%
\providecommand \selectlanguage [0]{\@gobble}%
\providecommand \bibinfo  [0]{\@secondoftwo}%
\providecommand \bibfield  [0]{\@secondoftwo}%
\providecommand \translation [1]{[#1]}%
\providecommand \BibitemOpen [0]{}%
\providecommand \bibitemStop [0]{}%
\providecommand \bibitemNoStop [0]{.\EOS\space}%
\providecommand \EOS [0]{\spacefactor3000\relax}%
\providecommand \BibitemShut  [1]{\csname bibitem#1\endcsname}%
\let\auto@bib@innerbib\@empty
\bibitem [{\citenamefont {M\o{}lhave}\ and\ \citenamefont
  {Drewsen}(2000)}]{Drewsen_2000_PR}%
  \BibitemOpen
  \bibfield  {author} {\bibinfo {author} {\bibfnamefont {K.}~\bibnamefont
  {M\o{}lhave}}\ and\ \bibinfo {author} {\bibfnamefont {M.}~\bibnamefont
  {Drewsen}},\ }\bibfield  {title} {\enquote {\bibinfo {title} {Formation of
  translationally cold {MgH}$^{+}$ and {MgD}$^{+}$ molecules in an ion trap},}\
  }\href@noop {} {\bibfield  {journal} {\bibinfo  {journal} {Phys. Rev. A}\
  }\textbf {\bibinfo {volume} {62}},\ \bibinfo {pages} {011401} (\bibinfo
  {year} {2000})}\BibitemShut {NoStop}%
\bibitem [{\citenamefont {Blythe}\ \emph {et~al.}(2005)\citenamefont {Blythe},
  \citenamefont {Roth}, \citenamefont {Fr\"{o}hlich}, \citenamefont {Wenz},\ and\
  \citenamefont {Schiller}}]{Blythe_2005_PR}%
  \BibitemOpen
  \bibfield  {author} {\bibinfo {author} {\bibfnamefont {P.}~\bibnamefont
  {Blythe}}, \bibinfo {author} {\bibfnamefont {B.}~\bibnamefont {Roth}},
  \bibinfo {author} {\bibfnamefont {U.}~\bibnamefont {Fr\"{o}hlich}}, \bibinfo
  {author} {\bibfnamefont {H.}~\bibnamefont {Wenz}}, \ and\ \bibinfo {author}
  {\bibfnamefont {S.}~\bibnamefont {Schiller}},\ }\bibfield  {title} {\enquote
  {\bibinfo {title} {Production of ultracold trapped molecular hydrogen
  ions},}\ }\href@noop {} {\bibfield  {journal} {\bibinfo  {journal} {Phys.
  Rev. Lett.}\ }\textbf {\bibinfo {volume} {95}},\ \bibinfo {pages} {183002}
  (\bibinfo {year} {2005})}\BibitemShut {NoStop}%
\bibitem [{\citenamefont {Baba}\ and\ \citenamefont
  {Waki}(2002)}]{Baba_2002_JCP}%
  \BibitemOpen
  \bibfield  {author} {\bibinfo {author} {\bibfnamefont {T.}~\bibnamefont
  {Baba}}\ and\ \bibinfo {author} {\bibfnamefont {I.}~\bibnamefont {Waki}},\
  }\bibfield  {title} {\enquote {\bibinfo {title} {Chemical reaction of
  sympathetically laser-cooled molecular ions},}\ }\href {\doibase
  http://dx.doi.org/10.1063/1.1431273} {\bibfield  {journal} {\bibinfo
  {journal} {J. Chem. Phys.}\ }\textbf {\bibinfo {volume} {116}},\ \bibinfo
  {pages} {1858--1861} (\bibinfo {year} {2002})}\BibitemShut {NoStop}%
\bibitem [{\citenamefont {Tong}\ \emph {et~al.}(2012)\citenamefont {Tong},
  \citenamefont {Nagy}, \citenamefont {Reyes}, \citenamefont {Germann},
  \citenamefont {Meuwly},\ and\ \citenamefont {Willitsch}}]{Tong_CPL_2012}%
  \BibitemOpen
  \bibfield  {author} {\bibinfo {author} {\bibfnamefont {X.}~\bibnamefont
  {Tong}}, \bibinfo {author} {\bibfnamefont {T.}~\bibnamefont {Nagy}}, \bibinfo
  {author} {\bibfnamefont {J.~Y.}\ \bibnamefont {Reyes}}, \bibinfo {author}
  {\bibfnamefont {M.}~\bibnamefont {Germann}}, \bibinfo {author} {\bibfnamefont
  {M.}~\bibnamefont {Meuwly}}, \ and\ \bibinfo {author} {\bibfnamefont
  {S.}~\bibnamefont {Willitsch}},\ }\bibfield  {title} {\enquote {\bibinfo
  {title} {State-selected ion-molecule reactions with {C}oulomb-crystallized
  molecular ions in traps},}\ }\href {\doibase
  http://dx.doi.org/10.1016/j.cplett.2012.06.042} {\bibfield  {journal}
  {\bibinfo  {journal} {Chem. Phys. Lett.}\ }\textbf {\bibinfo {volume}
  {547}},\ \bibinfo {pages} {1--8} (\bibinfo {year} {2012})}\BibitemShut
  {NoStop}%
\bibitem [{\citenamefont {Chang}\ \emph {et~al.}(2013)\citenamefont {Chang},
  \citenamefont {D{\l}ugo{\l}\c{e}cki}, \citenamefont {K\"{u}pper}, \citenamefont
  {R\"{o}sch}, \citenamefont {Wild},\ and\ \citenamefont
  {Willitsch}}]{Chang_Science_2013}%
  \BibitemOpen
  \bibfield  {author} {\bibinfo {author} {\bibfnamefont {Y.-P.}\ \bibnamefont
  {Chang}}, \bibinfo {author} {\bibfnamefont {K.}~\bibnamefont
  {D{\l}ugo{\l}\c{e}cki}}, \bibinfo {author} {\bibfnamefont {J.}~\bibnamefont
  {K\"{u}pper}}, \bibinfo {author} {\bibfnamefont {D.}~\bibnamefont {R\"{o}sch}},
  \bibinfo {author} {\bibfnamefont {D.}~\bibnamefont {Wild}}, \ and\ \bibinfo
  {author} {\bibfnamefont {S.}~\bibnamefont {Willitsch}},\ }\bibfield  {title}
  {\enquote {\bibinfo {title} {Specific chemical reactivities of spatially
  separated 3-aminophenol conformers with cold {Ca}$^+$ ions},}\ }\href
  {\doibase 10.1126/science.1242271} {\bibfield  {journal} {\bibinfo  {journal}
  {Scienc}\ }\textbf {\bibinfo {volume} {342}},\ \bibinfo {pages} {98--101}
  (\bibinfo {year} {2013})}\BibitemShut {NoStop}%
\bibitem [{\citenamefont {R\"{o}sch}\ \emph {et~al.}(2014)\citenamefont
  {R\"{o}sch}, \citenamefont {Willitsch}, \citenamefont {Chang},\ and\
  \citenamefont {K\"{u}pper}}]{Rosch_JCP_2014}%
  \BibitemOpen
  \bibfield  {author} {\bibinfo {author} {\bibfnamefont {D.}~\bibnamefont
  {R\"{o}sch}}, \bibinfo {author} {\bibfnamefont {S.}~\bibnamefont {Willitsch}},
  \bibinfo {author} {\bibfnamefont {Y.-P.}\ \bibnamefont {Chang}}, \ and\
  \bibinfo {author} {\bibfnamefont {J.}~\bibnamefont {K\"{u}pper}},\ }\bibfield
  {title} {\enquote {\bibinfo {title} {Chemical reactions of conformationally
  selected 3-aminophenol molecules in a beam with {C}oulomb-crystallized
  {Ca}$^+$ ions},}\ }\href {\doibase http://dx.doi.org/10.1063/1.4869100}
  {\bibfield  {journal} {\bibinfo  {journal} {J. Chem. Phys.}\ }\textbf
  {\bibinfo {volume} {140}},\ \bibinfo {eid} {124202} (\bibinfo {year}
  {2014})}\BibitemShut {NoStop}%
\bibitem [{\citenamefont {Douglas}, \citenamefont {Frank},\ and\ \citenamefont
  {Mao}(2005)}]{Douglas_MSR_2005}%
  \BibitemOpen
  \bibfield  {author} {\bibinfo {author} {\bibfnamefont {D.~J.}\ \bibnamefont
  {Douglas}}, \bibinfo {author} {\bibfnamefont {A.~J.}\ \bibnamefont {Frank}},
  \ and\ \bibinfo {author} {\bibfnamefont {D.}~\bibnamefont {Mao}},\ }\bibfield
   {title} {\enquote {\bibinfo {title} {Linear ion traps in mass
  spectrometry},}\ }\href@noop {} {\bibfield  {journal} {\bibinfo  {journal}
  {Mass Spectro. Rev.}\ }\textbf {\bibinfo {volume} {24}},\ \bibinfo {pages}
  {1--29} (\bibinfo {year} {2005})}\BibitemShut {NoStop}%
\bibitem [{\citenamefont {Leanhardt}\ \emph {et~al.}(2011)\citenamefont
  {Leanhardt}, \citenamefont {Bohn}, \citenamefont {Loh}, \citenamefont
  {Maletinsky}, \citenamefont {Meyer}, \citenamefont {Sinclair}, \citenamefont
  {Stutz},\ and\ \citenamefont {Cornell}}]{Leanhardt_JMS_2011}%
  \BibitemOpen
  \bibfield  {author} {\bibinfo {author} {\bibfnamefont {A.~E.}\ \bibnamefont
  {Leanhardt}}, \bibinfo {author} {\bibfnamefont {J.~L.}\ \bibnamefont {Bohn}},
  \bibinfo {author} {\bibfnamefont {H.}~\bibnamefont {Loh}}, \bibinfo {author}
  {\bibfnamefont {P.}~\bibnamefont {Maletinsky}}, \bibinfo {author}
  {\bibfnamefont {E.~R.}\ \bibnamefont {Meyer}}, \bibinfo {author}
  {\bibfnamefont {L.~C.}\ \bibnamefont {Sinclair}}, \bibinfo {author}
  {\bibfnamefont {R.~P.}\ \bibnamefont {Stutz}}, \ and\ \bibinfo {author}
  {\bibfnamefont {E.~A.}\ \bibnamefont {Cornell}},\ }\bibfield  {title}
  {\enquote {\bibinfo {title} {High-resolution spectroscopy on trapped
  molecular ions in rotating electric fields: A new approach for measuring the
  electron electric dipole moment},}\ }\href {\doibase
  http://dx.doi.org/10.1016/j.jms.2011.06.007} {\bibfield  {journal} {\bibinfo
  {journal} {J. Mol. Spectrosc.}\ }\textbf {\bibinfo {volume} {270}},\ \bibinfo
  {pages} {1--25} (\bibinfo {year} {2011})}\BibitemShut {NoStop}%
\bibitem [{\citenamefont {Loh}\ \emph {et~al.}(2013)\citenamefont {Loh},
  \citenamefont {Cossel}, \citenamefont {Grau}, \citenamefont {Ni},
  \citenamefont {Meyer}, \citenamefont {Bohn}, \citenamefont {Ye},\ and\
  \citenamefont {Cornell}}]{Loh_Science_2013}%
  \BibitemOpen
  \bibfield  {author} {\bibinfo {author} {\bibfnamefont {H.}~\bibnamefont
  {Loh}}, \bibinfo {author} {\bibfnamefont {K.~C.}\ \bibnamefont {Cossel}},
  \bibinfo {author} {\bibfnamefont {M.~C.}\ \bibnamefont {Grau}}, \bibinfo
  {author} {\bibfnamefont {K.-K.}\ \bibnamefont {Ni}}, \bibinfo {author}
  {\bibfnamefont {E.~R.}\ \bibnamefont {Meyer}}, \bibinfo {author}
  {\bibfnamefont {J.~L.}\ \bibnamefont {Bohn}}, \bibinfo {author}
  {\bibfnamefont {J.}~\bibnamefont {Ye}}, \ and\ \bibinfo {author}
  {\bibfnamefont {E.~A.}\ \bibnamefont {Cornell}},\ }\bibfield  {title}
  {\enquote {\bibinfo {title} {Precision spectroscopy of polarized molecules in
  an ion trap},}\ }\href {\doibase 10.1126/science.1243683} {\bibfield
  {journal} {\bibinfo  {journal} {Science}\ }\textbf {\bibinfo {volume}
  {342}},\ \bibinfo {pages} {1220--1222} (\bibinfo {year} {2013})}\BibitemShut
  {NoStop}%
\bibitem [{\citenamefont {Weidinger}\ and\ \citenamefont
  {Gruebele}(2008)}]{Weidinger_CP_2008}%
  \BibitemOpen
  \bibfield  {author} {\bibinfo {author} {\bibfnamefont {D.}~\bibnamefont
  {Weidinger}}\ and\ \bibinfo {author} {\bibfnamefont {M.}~\bibnamefont
  {Gruebele}},\ }\bibfield  {title} {\enquote {\bibinfo {title} {Simulations of
  quantum computation with a molecular ion},}\ }\href@noop {} {\bibfield
  {journal} {\bibinfo  {journal} {Chem. Phys.}\ }\textbf {\bibinfo {volume}
  {350}},\ \bibinfo {pages} {139--144} (\bibinfo {year} {2008})}\BibitemShut
  {NoStop}%
\bibitem [{\citenamefont {Buluta}\ \emph {et~al.}(2008)\citenamefont {Buluta},
  \citenamefont {Kitaoka}, \citenamefont {Georgescu},\ and\ \citenamefont
  {Hasegawa}}]{Buluta_2008_APS}%
  \BibitemOpen
  \bibfield  {author} {\bibinfo {author} {\bibfnamefont {I.~M.}\ \bibnamefont
  {Buluta}}, \bibinfo {author} {\bibfnamefont {M.}~\bibnamefont {Kitaoka}},
  \bibinfo {author} {\bibfnamefont {S.}~\bibnamefont {Georgescu}}, \ and\
  \bibinfo {author} {\bibfnamefont {S.}~\bibnamefont {Hasegawa}},\ }\bibfield
  {title} {\enquote {\bibinfo {title} {Investigation of planar {C}oulomb
  crystals for quantum simulation and computation},}\ }\href@noop {} {\bibfield
   {journal} {\bibinfo  {journal} {Phys. Rev. A}\ }\textbf {\bibinfo {volume}
  {77}},\ \bibinfo {pages} {062320} (\bibinfo {year} {2008})}\BibitemShut
  {NoStop}%
\bibitem [{\citenamefont {Alonso}\ \emph {et~al.}(2013)\citenamefont {Alonso},
  \citenamefont {Leupold}, \citenamefont {Keitch},\ and\ \citenamefont
  {Home}}]{Alonso_NJP_2013}%
  \BibitemOpen
  \bibfield  {author} {\bibinfo {author} {\bibfnamefont {J.}~\bibnamefont
  {Alonso}}, \bibinfo {author} {\bibfnamefont {F.~M.}\ \bibnamefont {Leupold}},
  \bibinfo {author} {\bibfnamefont {B.~C.}\ \bibnamefont {Keitch}}, \ and\
  \bibinfo {author} {\bibfnamefont {J.~P.}\ \bibnamefont {Home}},\ }\bibfield
  {title} {\enquote {\bibinfo {title} {Quantum control of the motional states
  of trapped ions through fast switching of trapping potentials},}\ }\href@noop
  {} {\bibfield  {journal} {\bibinfo  {journal} {New J. Phys.}\ }\textbf
  {\bibinfo {volume} {15}},\ \bibinfo {pages} {023001} (\bibinfo {year}
  {2013})}\BibitemShut {NoStop}%
\bibitem [{\citenamefont {Amniat-Talab}, \citenamefont {Saadati-Niari},\ and\
  \citenamefont {Gu\'{e}rin}(2012)}]{Amniat-Talab_EPJD_2012}%
  \BibitemOpen
  \bibfield  {author} {\bibinfo {author} {\bibfnamefont {M.}~\bibnamefont
  {Amniat-Talab}}, \bibinfo {author} {\bibfnamefont {M.}~\bibnamefont
  {Saadati-Niari}}, \ and\ \bibinfo {author} {\bibfnamefont {S.}~\bibnamefont
  {Gu\'{e}rin}},\ }\bibfield  {title} {\enquote {\bibinfo {title} {Quantum state
  engineering in ion-traps via adiabatic passage},}\ }\href@noop {} {\bibfield
  {journal} {\bibinfo  {journal} {Eur. Phys. J. D}\ }\textbf {\bibinfo {volume}
  {66}},\ \bibinfo {pages} {1--9} (\bibinfo {year} {2012})}\BibitemShut
  {NoStop}%
\bibitem [{\citenamefont {Leibfried}(2012)}]{Leibfried_NJP_2012}%
  \BibitemOpen
  \bibfield  {author} {\bibinfo {author} {\bibfnamefont {D.}~\bibnamefont
  {Leibfried}},\ }\bibfield  {title} {\enquote {\bibinfo {title} {Quantum state
  preparation and control of single molecular ions},}\ }\href
  {http://stacks.iop.org/1367-2630/14/i=2/a=023029} {\bibfield  {journal}
  {\bibinfo  {journal} {New J. Phys.}\ }\textbf {\bibinfo {volume} {14}},\
  \bibinfo {pages} {023029} (\bibinfo {year} {2012})}\BibitemShut {NoStop}%
\bibitem [{\citenamefont {Shi}\ \emph {et~al.}(2013)\citenamefont {Shi},
  \citenamefont {Herskind}, \citenamefont {Drewsen},\ and\ \citenamefont
  {Chuang}}]{Shi_NJP_2013}%
  \BibitemOpen
  \bibfield  {author} {\bibinfo {author} {\bibfnamefont {M.}~\bibnamefont
  {Shi}}, \bibinfo {author} {\bibfnamefont {P.~F.}\ \bibnamefont {Herskind}},
  \bibinfo {author} {\bibfnamefont {M.}~\bibnamefont {Drewsen}}, \ and\
  \bibinfo {author} {\bibfnamefont {I.~L.}\ \bibnamefont {Chuang}},\ }\bibfield
   {title} {\enquote {\bibinfo {title} {Microwave quantum logic spectroscopy
  and control of molecular ions},}\ }\href
  {http://stacks.iop.org/1367-2630/15/i=11/a=113019} {\bibfield  {journal}
  {\bibinfo  {journal} {New J. Phys.}\ }\textbf {\bibinfo {volume} {15}},\
  \bibinfo {pages} {113019} (\bibinfo {year} {2013})}\BibitemShut {NoStop}%
\bibitem [{\citenamefont {Paul}(1990)}]{Paul_RMP_1990}%
  \BibitemOpen
  \bibfield  {author} {\bibinfo {author} {\bibfnamefont {W.}~\bibnamefont
  {Paul}},\ }\bibfield  {title} {\enquote {\bibinfo {title} {Electromagnetic
  traps for charged and neutral particles},}\ }\href {\doibase
  10.1103/RevModPhys.62.531} {\bibfield  {journal} {\bibinfo  {journal} {Rev.
  Mod. Phys.}\ }\textbf {\bibinfo {volume} {62}},\ \bibinfo {pages} {531--540}
  (\bibinfo {year} {1990})}\BibitemShut {NoStop}%
\bibitem [{\citenamefont {Prestage}, \citenamefont {Dick},\ and\ \citenamefont
  {Maleki}(1989)}]{Prestage_JAP_1989}%
  \BibitemOpen
  \bibfield  {author} {\bibinfo {author} {\bibfnamefont {J.~D.}\ \bibnamefont
  {Prestage}}, \bibinfo {author} {\bibfnamefont {G.~J.}\ \bibnamefont {Dick}},
  \ and\ \bibinfo {author} {\bibfnamefont {L.}~\bibnamefont {Maleki}},\
  }\bibfield  {title} {\enquote {\bibinfo {title} {New ion trap for frequency
  standard applications},}\ }\href {\doibase
  http://dx.doi.org/10.1063/1.343486} {\bibfield  {journal} {\bibinfo
  {journal} {J. Appl. Phys.}\ }\textbf {\bibinfo {volume} {66}},\ \bibinfo
  {pages} {1013--1017} (\bibinfo {year} {1989})}\BibitemShut {NoStop}%
\bibitem [{\citenamefont {Raizen}\ \emph
  {et~al.}(1992{\natexlab{a}})\citenamefont {Raizen}, \citenamefont {Gilligan},
  \citenamefont {Bergquist}, \citenamefont {Itano},\ and\ \citenamefont
  {Wineland}}]{Raizen_PRA_1992}%
  \BibitemOpen
  \bibfield  {author} {\bibinfo {author} {\bibfnamefont {M.~G.}\ \bibnamefont
  {Raizen}}, \bibinfo {author} {\bibfnamefont {J.~M.}\ \bibnamefont
  {Gilligan}}, \bibinfo {author} {\bibfnamefont {J.~C.}\ \bibnamefont
  {Bergquist}}, \bibinfo {author} {\bibfnamefont {W.~M.}\ \bibnamefont
  {Itano}}, \ and\ \bibinfo {author} {\bibfnamefont {D.~J.}\ \bibnamefont
  {Wineland}},\ }\bibfield  {title} {\enquote {\bibinfo {title} {Ionic crystals
  in a linear {P}aul trap},}\ }\href {\doibase 10.1103/PhysRevA.45.6493}
  {\bibfield  {journal} {\bibinfo  {journal} {Phys. Rev. A}\ }\textbf {\bibinfo
  {volume} {45}},\ \bibinfo {pages} {6493--6501} (\bibinfo {year}
  {1992}{\natexlab{a}})}\BibitemShut {NoStop}%
\bibitem [{\citenamefont {Raizen}\ \emph
  {et~al.}(1992{\natexlab{b}})\citenamefont {Raizen}, \citenamefont {Gilligan},
  \citenamefont {Bergquist}, \citenamefont {Itano},\ and\ \citenamefont
  {Wineland}}]{Raizen_JMO_1992}%
  \BibitemOpen
  \bibfield  {author} {\bibinfo {author} {\bibfnamefont {M.~G.}\ \bibnamefont
  {Raizen}}, \bibinfo {author} {\bibfnamefont {J.~M.}\ \bibnamefont
  {Gilligan}}, \bibinfo {author} {\bibfnamefont {J.~C.}\ \bibnamefont
  {Bergquist}}, \bibinfo {author} {\bibfnamefont {W.~M.}\ \bibnamefont
  {Itano}}, \ and\ \bibinfo {author} {\bibfnamefont {D.~J.}\ \bibnamefont
  {Wineland}},\ }\bibfield  {title} {\enquote {\bibinfo {title} {Linear trap
  for high-accuracy spectroscopy of stored ions},}\ }\href@noop {} {\bibfield
  {journal} {\bibinfo  {journal} {J. Mod. Opt.}\ }\textbf {\bibinfo {volume}
  {39}},\ \bibinfo {pages} {233--242} (\bibinfo {year}
  {1992}{\natexlab{b}})}\BibitemShut {NoStop}%
\bibitem [{\citenamefont {Brki\'{c}}\ \emph {et~al.}(2006)\citenamefont
  {Brki\'{c}}, \citenamefont {Taylor}, \citenamefont {Ralph},\ and\
  \citenamefont {France}}]{Brkic_PRA_2006}%
  \BibitemOpen
  \bibfield  {author} {\bibinfo {author} {\bibfnamefont {B.}~\bibnamefont
  {Brki\'{c}}}, \bibinfo {author} {\bibfnamefont {S.}~\bibnamefont {Taylor}},
  \bibinfo {author} {\bibfnamefont {J.~F.}\ \bibnamefont {Ralph}}, \ and\
  \bibinfo {author} {\bibfnamefont {N.}~\bibnamefont {France}},\ }\bibfield
  {title} {\enquote {\bibinfo {title} {High-fidelity simulations of ion
  trajectories in miniature ion traps using the boundary-element method},}\
  }\href@noop {} {\bibfield  {journal} {\bibinfo  {journal} {Phys. Rev. A}\
  }\textbf {\bibinfo {volume} {73}},\ \bibinfo {pages} {012326} (\bibinfo
  {year} {2006})}\BibitemShut {NoStop}%
\bibitem [{\citenamefont {Drewsen}\ \emph {et~al.}(2003)\citenamefont
  {Drewsen}, \citenamefont {Jensen}, \citenamefont {Lindballe}, \citenamefont
  {Nissen}, \citenamefont {Martinussen}, \citenamefont {Mortensen},
  \citenamefont {Staanum},\ and\ \citenamefont {Voigt}}]{Drewsen_2003_MS}%
  \BibitemOpen
  \bibfield  {author} {\bibinfo {author} {\bibfnamefont {M.}~\bibnamefont
  {Drewsen}}, \bibinfo {author} {\bibfnamefont {I.}~\bibnamefont {Jensen}},
  \bibinfo {author} {\bibfnamefont {J.}~\bibnamefont {Lindballe}}, \bibinfo
  {author} {\bibfnamefont {N.}~\bibnamefont {Nissen}}, \bibinfo {author}
  {\bibfnamefont {R.}~\bibnamefont {Martinussen}}, \bibinfo {author}
  {\bibfnamefont {A.}~\bibnamefont {Mortensen}}, \bibinfo {author}
  {\bibfnamefont {P.}~\bibnamefont {Staanum}}, \ and\ \bibinfo {author}
  {\bibfnamefont {D.}~\bibnamefont {Voigt}},\ }\bibfield  {title} {\enquote
  {\bibinfo {title} {Ion {C}oulomb crystals: a tool for studying ion
  processes},}\ }\href@noop {} {\bibfield  {journal} {\bibinfo  {journal} {Int.
  J. Mass Spectrom.}\ }\textbf {\bibinfo {volume} {229}},\ \bibinfo {pages}
  {83--91} (\bibinfo {year} {2003})}\BibitemShut {NoStop}%
\bibitem [{\citenamefont {Porras}\ and\ \citenamefont
  {Cirac}(2006)}]{Porras_2006_PR}%
  \BibitemOpen
  \bibfield  {author} {\bibinfo {author} {\bibfnamefont {D.}~\bibnamefont
  {Porras}}\ and\ \bibinfo {author} {\bibfnamefont {J.~I.}\ \bibnamefont
  {Cirac}},\ }\bibfield  {title} {\enquote {\bibinfo {title} {Quantum
  manipulation of trapped ions in two dimensional {C}oulomb crystals},}\
  }\href@noop {} {\bibfield  {journal} {\bibinfo  {journal} {Phys. Rev. Lett.}\
  }\textbf {\bibinfo {volume} {96}},\ \bibinfo {pages} {250501} (\bibinfo
  {year} {2006})}\BibitemShut {NoStop}%
\bibitem [{\citenamefont {Willitsch}(2012)}]{Willitsch_IRPC_2012}%
  \BibitemOpen
  \bibfield  {author} {\bibinfo {author} {\bibfnamefont {S.}~\bibnamefont
  {Willitsch}},\ }\bibfield  {title} {\enquote {\bibinfo {title}
  {{C}oulomb-crystallised molecular ions in traps: methods, applications,
  prospects},}\ }\href {\doibase 10.1080/0144235X.2012.667221} {\bibfield
  {journal} {\bibinfo  {journal} {Int. Rev. Phys. Chem.}\ }\textbf {\bibinfo
  {volume} {31}},\ \bibinfo {pages} {175--199} (\bibinfo {year}
  {2012})}\BibitemShut {NoStop}%
\bibitem [{\citenamefont {Vogelius}, \citenamefont {Madsen},\ and\
  \citenamefont {Drewsen}(2006)}]{Vogelius_JPB_2006}%
  \BibitemOpen
  \bibfield  {author} {\bibinfo {author} {\bibfnamefont {I.~S.}\ \bibnamefont
  {Vogelius}}, \bibinfo {author} {\bibfnamefont {L.~B.}\ \bibnamefont
  {Madsen}}, \ and\ \bibinfo {author} {\bibfnamefont {M.}~\bibnamefont
  {Drewsen}},\ }\bibfield  {title} {\enquote {\bibinfo {title} {Rotational
  cooling of molecular ions through laser-induced coupling to the collective
  modes of a two-ion {C}oulomb crystal},}\ }\href
  {http://stacks.iop.org/0953-4075/39/i=19/a=S32} {\bibfield  {journal}
  {\bibinfo  {journal} {J. Phys. B}\ }\textbf {\bibinfo {volume} {39}},\
  \bibinfo {pages} {S1267--S1280} (\bibinfo {year} {2006})}\BibitemShut
  {NoStop}%
\bibitem [{\citenamefont {Lazarou}, \citenamefont {Keller},\ and\ \citenamefont
  {Garraway}(2010)}]{Lazarou_PRA_2010}%
  \BibitemOpen
  \bibfield  {author} {\bibinfo {author} {\bibfnamefont {C.}~\bibnamefont
  {Lazarou}}, \bibinfo {author} {\bibfnamefont {M.}~\bibnamefont {Keller}}, \
  and\ \bibinfo {author} {\bibfnamefont {B.~M.}\ \bibnamefont {Garraway}},\
  }\bibfield  {title} {\enquote {\bibinfo {title} {Molecular heat pump for
  rotational states},}\ }\href {\doibase 10.1103/PhysRevA.81.013418} {\bibfield
   {journal} {\bibinfo  {journal} {Phys. Rev. A}\ }\textbf {\bibinfo {volume}
  {81}},\ \bibinfo {pages} {013418} (\bibinfo {year} {2010})}\BibitemShut
  {NoStop}%
\bibitem [{\citenamefont {Berglund}, \citenamefont {Drewsen},\ and\
  \citenamefont {Koch}(2015)}]{Berglund_2015_NJP}%
  \BibitemOpen
  \bibfield  {author} {\bibinfo {author} {\bibfnamefont {J.~M.}\ \bibnamefont
  {Berglund}}, \bibinfo {author} {\bibfnamefont {M.}~\bibnamefont {Drewsen}}, \
  and\ \bibinfo {author} {\bibfnamefont {C.~P.}\ \bibnamefont {Koch}},\
  }\bibfield  {title} {\enquote {\bibinfo {title} {Femtosecond wavepacket
  interferometry using the rotational dynamics of a trapped cold molecular
  ion},}\ }\href@noop {} {\bibfield  {journal} {\bibinfo  {journal} {New J.
  Phys.}\ }\textbf {\bibinfo {volume} {17}},\ \bibinfo {pages} {025007}
  (\bibinfo {year} {2015})}\BibitemShut {NoStop}%
\bibitem [{\citenamefont {Evans}(1977)}]{evans1977representatation}%
  \BibitemOpen
  \bibfield  {author} {\bibinfo {author} {\bibfnamefont {D.~J.}\ \bibnamefont
  {Evans}},\ }\bibfield  {title} {\enquote {\bibinfo {title} {On the
  representation of orientation space},}\ }\href@noop {} {\bibfield  {journal}
  {\bibinfo  {journal} {Mol. Phys.}\ }\textbf {\bibinfo {volume} {34}},\
  \bibinfo {pages} {317--325} (\bibinfo {year} {1977})}\BibitemShut {NoStop}%
\bibitem [{\citenamefont {Major}, \citenamefont {Gheorghe},\ and\ \citenamefont
  {Werth}(2005)}]{Major_book_2005}%
  \BibitemOpen
  \bibfield  {author} {\bibinfo {author} {\bibfnamefont {F.~G.}\ \bibnamefont
  {Major}}, \bibinfo {author} {\bibfnamefont {V.~N.}\ \bibnamefont {Gheorghe}},
  \ and\ \bibinfo {author} {\bibfnamefont {G.}~\bibnamefont {Werth}},\
  }\href@noop {} {\emph {\bibinfo {title} {Charged Particle Traps: Physics and
  Techniques of Charged Particle Field Confinement}}}\ (\bibinfo  {publisher}
  {Springer},\ \bibinfo {address} {Berlin},\ \bibinfo {year}
  {2005})\BibitemShut {NoStop}%
\bibitem [{\citenamefont {Ghosh}(1995)}]{Ghosh_book_1995}%
  \BibitemOpen
  \bibfield  {author} {\bibinfo {author} {\bibfnamefont {P.~K.}\ \bibnamefont
  {Ghosh}},\ }\href@noop {} {\emph {\bibinfo {title} {Ion Traps}}}\ (\bibinfo
  {publisher} {Oxford University Press},\ \bibinfo {address} {Oxford},\
  \bibinfo {year} {1995})\BibitemShut {NoStop}%
\bibitem [{\citenamefont {Hashemloo}, \citenamefont {Dion},\ and\ \citenamefont
  {Rahali}(2016)}]{Hashemloo_IJMPC_2016}%
  \BibitemOpen
  \bibfield  {author} {\bibinfo {author} {\bibfnamefont {A.}~\bibnamefont
  {Hashemloo}}, \bibinfo {author} {\bibfnamefont {C.~M.}\ \bibnamefont {Dion}},
  \ and\ \bibinfo {author} {\bibfnamefont {G.}~\bibnamefont {Rahali}},\
  }\bibfield  {title} {\enquote {\bibinfo {title} {Wave-packet dynamics of an
  atomic ion in a {P}aul trap: Approximations and stability},}\ }\href
  {\doibase 10.1142/S0129183116500145} {\bibfield  {journal} {\bibinfo
  {journal} {Int. J. Mod. Phys. C}\ }\textbf {\bibinfo {volume} {27}},\
  \bibinfo {pages} {1650014} (\bibinfo {year} {2016})}\BibitemShut {NoStop}%
\bibitem [{\citenamefont {Wolf}(2010)}]{Wolf_NIST_2010}%
  \BibitemOpen
  \bibfield  {author} {\bibinfo {author} {\bibfnamefont {G.}~\bibnamefont
  {Wolf}},\ }\bibfield  {title} {\enquote {\bibinfo {title} {{M}athieu
  functions and {H}ill's equation},}\ }in\ \href@noop {} {\emph {\bibinfo
  {booktitle} {{NIST} Handbook of Mathematical Functions}}},\ \bibinfo {editor}
  {edited by\ \bibinfo {editor} {\bibfnamefont {F.~W.~J.}\ \bibnamefont
  {Oliver}}, \bibinfo {editor} {\bibfnamefont {D.~W.}\ \bibnamefont {Lozier}},
  \bibinfo {editor} {\bibfnamefont {R.~F.}\ \bibnamefont {Boisvert}}, \ and\
  \bibinfo {editor} {\bibfnamefont {C.~W.}\ \bibnamefont {Clark}}}\ (\bibinfo
  {publisher} {Cambridge University Press},\ \bibinfo {address} {Cambridge},\
  \bibinfo {year} {2010})\ Chap.~\bibinfo {chapter} {28}, pp.\ \bibinfo {pages}
  {651--681}\BibitemShut {NoStop}%
\bibitem [{\citenamefont {Cheung}\ and\ \citenamefont
  {Powles}(1975)}]{Cheung_1}%
  \BibitemOpen
  \bibfield  {author} {\bibinfo {author} {\bibfnamefont {P.~S.~Y.}\
  \bibnamefont {Cheung}}\ and\ \bibinfo {author} {\bibfnamefont {J.~G.}\
  \bibnamefont {Powles}},\ }\bibfield  {title} {\enquote {\bibinfo {title} {The
  properties of liquid nitrogen},}\ }\href@noop {} {\bibfield  {journal}
  {\bibinfo  {journal} {Mol. Phys.}\ }\textbf {\bibinfo {volume} {30}},\
  \bibinfo {pages} {921--949} (\bibinfo {year} {1975})}\BibitemShut {NoStop}%
\bibitem [{\citenamefont {Cheung}(1976)}]{Cheung_2}%
  \BibitemOpen
  \bibfield  {author} {\bibinfo {author} {\bibfnamefont {P.~S.~Y.}\
  \bibnamefont {Cheung}},\ }\bibfield  {title} {\enquote {\bibinfo {title} {On
  the efficient evaluation of torques and forces for anisotropic potentials in
  computer simulation of liquids composed of linear molecules},}\ }\href@noop
  {} {\bibfield  {journal} {\bibinfo  {journal} {Chem. Phys. Lett.}\ }\textbf
  {\bibinfo {volume} {40}},\ \bibinfo {pages} {19 -- 22} (\bibinfo {year}
  {1976})}\BibitemShut {NoStop}%
\bibitem [{\citenamefont {Evans}\ and\ \citenamefont
  {Murad}(1977)}]{evans1977singularity}%
  \BibitemOpen
  \bibfield  {author} {\bibinfo {author} {\bibfnamefont {D.~J.}\ \bibnamefont
  {Evans}}\ and\ \bibinfo {author} {\bibfnamefont {S.}~\bibnamefont {Murad}},\
  }\bibfield  {title} {\enquote {\bibinfo {title} {Singularity free algorithm
  for molecular dynamics simulation of rigid polyatomics},}\ }\href@noop {}
  {\bibfield  {journal} {\bibinfo  {journal} {Mol. Phys.}\ }\textbf {\bibinfo
  {volume} {34}},\ \bibinfo {pages} {327--331} (\bibinfo {year}
  {1977})}\BibitemShut {NoStop}%
\bibitem [{\citenamefont {Goldstein}, \citenamefont {Poole},\ and\
  \citenamefont {Safko}(2002)}]{goldstein2002classical}%
  \BibitemOpen
  \bibfield  {author} {\bibinfo {author} {\bibfnamefont {H.}~\bibnamefont
  {Goldstein}}, \bibinfo {author} {\bibfnamefont {C.~P.}\ \bibnamefont
  {Poole}}, \ and\ \bibinfo {author} {\bibfnamefont {J.~L.}\ \bibnamefont
  {Safko}},\ }\href@noop {} {\emph {\bibinfo {title} {Classical Mechanics}}},\
  \bibinfo {edition} {3rd}\ ed.\ (\bibinfo  {publisher} {Addison Wesley},\
  \bibinfo {address} {San Francisco},\ \bibinfo {year} {2002})\BibitemShut
  {NoStop}%
\bibitem [{\citenamefont {Combescure}(1986)}]{Combescure_AIHPA_1986}%
  \BibitemOpen
  \bibfield  {author} {\bibinfo {author} {\bibfnamefont {M.}~\bibnamefont
  {Combescure}},\ }\bibfield  {title} {\enquote {\bibinfo {title} {A quantum
  particle in a quadrupole radio-frequency trap},}\ }\href@noop {} {\bibfield
  {journal} {\bibinfo  {journal} {Ann. Inst. Henri Poincar\'{e}, A}\ }\textbf
  {\bibinfo {volume} {44}},\ \bibinfo {pages} {293--314} (\bibinfo {year}
  {1986})}\BibitemShut {NoStop}%
\bibitem [{\citenamefont {Brown}(1991)}]{Brown_PRL_1991}%
  \BibitemOpen
  \bibfield  {author} {\bibinfo {author} {\bibfnamefont {L.~S.}\ \bibnamefont
  {Brown}},\ }\bibfield  {title} {\enquote {\bibinfo {title} {Quantum motion in
  a {P}aul trap},}\ }\href {\doibase 10.1103/PhysRevLett.66.527} {\bibfield
  {journal} {\bibinfo  {journal} {Phys. Rev. Lett.}\ }\textbf {\bibinfo
  {volume} {66}},\ \bibinfo {pages} {527--529} (\bibinfo {year}
  {1991})}\BibitemShut {NoStop}%
\bibitem [{\citenamefont {Glauber}(1992)}]{Glauber_1992}%
  \BibitemOpen
  \bibfield  {author} {\bibinfo {author} {\bibfnamefont {R.~J.}\ \bibnamefont
  {Glauber}},\ }\bibfield  {title} {\enquote {\bibinfo {title} {The quantum
  mechanics of trapped wave packets},}\ }in\ \href@noop {} {\emph {\bibinfo
  {booktitle} {Laser Manipulation of Atoms and Ions}}},\ \bibinfo {editor}
  {edited by\ \bibinfo {editor} {\bibfnamefont {E.}~\bibnamefont {Arimondo}},
  \bibinfo {editor} {\bibfnamefont {W.~D.}\ \bibnamefont {Phillips}}, \ and\
  \bibinfo {editor} {\bibfnamefont {F.}~\bibnamefont {Strumia}}}\ (\bibinfo
  {publisher} {North-Holland},\ \bibinfo {address} {Amsterdam},\ \bibinfo
  {year} {1992})\ pp.\ \bibinfo {pages} {643--660}\BibitemShut {NoStop}%
\bibitem [{\citenamefont {Varshalovich}, \citenamefont {Moskalev},\ and\
  \citenamefont {Khersonski{\u\i}}(1988)}]{varshalovich1988quantum}%
  \BibitemOpen
  \bibfield  {author} {\bibinfo {author} {\bibfnamefont {D.~A.}\ \bibnamefont
  {Varshalovich}}, \bibinfo {author} {\bibfnamefont {A.~N.}\ \bibnamefont
  {Moskalev}}, \ and\ \bibinfo {author} {\bibfnamefont {V.~K.}\ \bibnamefont
  {Khersonski{\u\i}}},\ }\href@noop {} {\emph {\bibinfo {title} {Quantum Theory
  of Angular Momentum}}}\ (\bibinfo  {publisher} {World Scientific},\ \bibinfo
  {address} {Singapore},\ \bibinfo {year} {1988})\BibitemShut {NoStop}%
\bibitem [{\citenamefont {Press}\ \emph {et~al.}(1992)\citenamefont {Press},
  \citenamefont {Teukolsky}, \citenamefont {Vetterling},\ and\ \citenamefont
  {Flannery}}]{pressC92}%
  \BibitemOpen
  \bibfield  {author} {\bibinfo {author} {\bibfnamefont {W.~H.}\ \bibnamefont
  {Press}}, \bibinfo {author} {\bibfnamefont {S.~A.}\ \bibnamefont
  {Teukolsky}}, \bibinfo {author} {\bibfnamefont {W.~T.}\ \bibnamefont
  {Vetterling}}, \ and\ \bibinfo {author} {\bibfnamefont {B.~P.}\ \bibnamefont
  {Flannery}},\ }\href@noop {} {\emph {\bibinfo {title} {Numerical Recipes in
  {C}}}},\ \bibinfo {edition} {2nd}\ ed.\ (\bibinfo  {publisher} {Cambridge
  University Press},\ \bibinfo {address} {Cambridge},\ \bibinfo {year}
  {1992})\BibitemShut {NoStop}%
\bibitem [{\citenamefont {Galassi~\emph{et al.}}(2009)}]{gsl}%
  \BibitemOpen
  \bibfield  {author} {\bibinfo {author} {\bibfnamefont {M.}~\bibnamefont
  {Galassi~\emph{et al.}}},\ }\href@noop {} {\emph {\bibinfo {title} {GNU
  Scientific Library Reference Manual}}},\ \bibinfo {edition} {3rd}\ ed.\
  (\bibinfo  {publisher} {Network Theory},\ \bibinfo {address} {Bristol},\
  \bibinfo {year} {2009})\BibitemShut {NoStop}%
\bibitem [{\citenamefont {Bertelsen}, \citenamefont {J\o{}rgensen},\ and\
  \citenamefont {Drewsen}(2006)}]{Drewsen_1}%
  \BibitemOpen
  \bibfield  {author} {\bibinfo {author} {\bibfnamefont {A.}~\bibnamefont
  {Bertelsen}}, \bibinfo {author} {\bibfnamefont {S.}~\bibnamefont
  {J\o{}rgensen}}, \ and\ \bibinfo {author} {\bibfnamefont {M.}~\bibnamefont
  {Drewsen}},\ }\bibfield  {title} {\enquote {\bibinfo {title} {The rotational
  temperature of polar molecular ions in {C}oulomb crystals},}\ }\href@noop {}
  {\bibfield  {journal} {\bibinfo  {journal} {J. Phys. B}\ }\textbf {\bibinfo
  {volume} {39}},\ \bibinfo {pages} {L83} (\bibinfo {year} {2006})}\BibitemShut
  {NoStop}%
\bibitem [{\citenamefont {Abe}\ \emph {et~al.}(2010)\citenamefont {Abe},
  \citenamefont {Kajita}, \citenamefont {Hada},\ and\ \citenamefont
  {Moriwaki}}]{Japan}%
  \BibitemOpen
  \bibfield  {author} {\bibinfo {author} {\bibfnamefont {M.}~\bibnamefont
  {Abe}}, \bibinfo {author} {\bibfnamefont {M.}~\bibnamefont {Kajita}},
  \bibinfo {author} {\bibfnamefont {M.}~\bibnamefont {Hada}}, \ and\ \bibinfo
  {author} {\bibfnamefont {Y.}~\bibnamefont {Moriwaki}},\ }\bibfield  {title}
  {\enquote {\bibinfo {title} {Ab initio study on vibrational dipole moments of
  {XH}$^+$ molecular ions: {X} = $^{24}${Mg}, $^{40}${Ca}, $^{64}${Zn},
  $^{88}${Sr}, $^{114}${Cd}, $^{138}${Ba}, $^{174}${Yb} and $^{202}${Hg}},}\
  }\href@noop {} {\bibfield  {journal} {\bibinfo  {journal} {J. Phys. B}\
  }\textbf {\bibinfo {volume} {43}},\ \bibinfo {pages} {245102} (\bibinfo
  {year} {2010})}\BibitemShut {NoStop}%
\bibitem [{\citenamefont {H\o{}jbjerre}\ \emph {et~al.}(2009)\citenamefont
  {H\o{}jbjerre}, \citenamefont {Hansen}, \citenamefont {Skyt}, \citenamefont
  {Staanum},\ and\ \citenamefont {Drewsen}}]{Drewsen_2}%
  \BibitemOpen
  \bibfield  {author} {\bibinfo {author} {\bibfnamefont {K.}~\bibnamefont
  {H\o{}jbjerre}}, \bibinfo {author} {\bibfnamefont {A.~K.}\ \bibnamefont
  {Hansen}}, \bibinfo {author} {\bibfnamefont {P.~S.}\ \bibnamefont {Skyt}},
  \bibinfo {author} {\bibfnamefont {P.~F.}\ \bibnamefont {Staanum}}, \ and\
  \bibinfo {author} {\bibfnamefont {M.}~\bibnamefont {Drewsen}},\ }\bibfield
  {title} {\enquote {\bibinfo {title} {Rotational state resolved
  photodissociation spectroscopy of translationally and vibrationally cold
  {MgH}$^+$ ions: toward rotational cooling of molecular ions},}\ }\href@noop
  {} {\bibfield  {journal} {\bibinfo  {journal} {New J. Phys.}\ }\textbf
  {\bibinfo {volume} {11}},\ \bibinfo {pages} {055026} (\bibinfo {year}
  {2009})}\BibitemShut {NoStop}%
\bibitem [{\citenamefont {von Meyenn}(1970)}]{vonMeyenn_ZP_1970}%
  \BibitemOpen
  \bibfield  {author} {\bibinfo {author} {\bibfnamefont {K.}~\bibnamefont {von
  Meyenn}},\ }\bibfield  {title} {\enquote {\bibinfo
  {title} {Rotation von zweiatomigen {D}ipolmolek\"{u}len in starken elektrischen
  {F}eldern},}\ }\href {\doibase 10.1007/BF01392506} {\bibfield  {journal}
  {\bibinfo  {journal} {Z. Phys.}\ }\textbf {\bibinfo {volume} {231}},\
  \bibinfo {pages} {154--160} (\bibinfo {year} {1970})}\BibitemShut {NoStop}%
\bibitem [{\citenamefont {Friedrich}\ and\ \citenamefont
  {Herschbach}(1991)}]{Friedrich_ZPD_1991}%
  \BibitemOpen
  \bibfield  {author} {\bibinfo {author} {\bibfnamefont {B.}~\bibnamefont
  {Friedrich}}\ and\ \bibinfo {author} {\bibfnamefont {D.~R.}\ \bibnamefont
  {Herschbach}},\ }\bibfield  {title} {\enquote {\bibinfo {title} {On the
  possibility of orienting rotationally cooled polar molecules in an electric
  field},}\ }\href@noop {} {\bibfield  {journal} {\bibinfo  {journal} {Z. Phys.
  D}\ }\textbf {\bibinfo {volume} {18}},\ \bibinfo {pages} {153--161} (\bibinfo
  {year} {1991})}\BibitemShut {NoStop}%
\end{thebibliography}%

\end{document}